# A new application of the Fox-Wright functions: the coherent states formalism

## Dušan POPOV[a, b]

[a] University Politehnica Timisoara, Department of Physical Foundations of Engineering, B-dul Vasile Pârvan No. 2, 300223 Timisoara, Romania
[b] Serbian Academy of Nonlinear Sciences (SANS), Kneza Mihaila 36, Beograd-Stari Grad, Belgrade, Serbia
E-mail: dusan_popov@yahoo.co.uk
ORCID: https://orcid.org/0000-0003-3631-3247

## Abstract

In this paper we extend the applicability of Fox-Wright functions beyond mathematics, specifically in quantum physics. We focused our attention on a new application, on the connection between the Fox-Wright functions and the generalized coherent states formalism. We constructed the generalized coherent states in the Barut-Girardello manner, in which the Fox-Wright functions play the role of normalization functions, and we demonstrated that the Fox-Wright coherent states satisfy all general conditions imposed on the set of coherent states. In parallel, we examined the properties of both pure and mixed (thermal) Fox-Wright coherent states. All calculations were performed within the diagonal operators ordering technique (DOOT) using the Dirac's bra-ket formalism. Finally, we introduced some (specifically, integral) feedback elements that Fox-Wright coherent states induce in the theory of special functions, including a new integral representation of Fox-Wright functions.

## 1. Introduction

Exactly a century ago, Schrödinger posed the problem of finding quantum states that mimic the behavior of corresponding classical states [1]. Thus, 1926 became the year of birth of coherent states (CSs). In the following decades, researchers focused on the properties and applications of the CSs of the one-dimensional harmonic oscillator (HO-1D). Since the formalism of these CSs implied the involvement of the annihilation and creation boson operators $\hat{a}$ and $\hat{a}^+$, which satisfy the canonical commutation relation $\left[\hat{a}, \hat{a}^+\right] = 1$, these CSs received the name canonical coherent states (CCSs) [2], [3], [4].

At the same time, however, it became clear that CSs cannot be limited only to HO-1D, whose energy spectrum is linear with respect to the main quantum number $n$, and this formalism can also be extended to systems with non-linear energy spectrum. Thus, new types of CSs ("non-canonical" CSs) appeared that received different names: nonlinear coherent states (NCSs), q-deformed coherent states (qCSs) or generalized hypergeometric coherent states (GHCSs), although their essence, respectively their mathematical expression has the similar structure. In fact, all these types of CSs are generalized coherent states (GCSs).

In this paper we use the Dirac bra-ket notation for the Fock vectors, as well as for the coherent states.

Generally speaking, the CSs are a set of vectors $|z>$, labeled by a complex variable $z = |z| \exp(i\varphi)$, $|z| \le R_c$, where $R_c$ is the convergence radius of the sum of corresponding power series, i.e. of their normalization function, and $0 \le \varphi \le 2\pi$. Any set of CSs $|z>$ can be expanded with respect to an orthogonal basis vectors, e.g. the Fock vectors, $\{|n>, \ n = 0, 1, 2, ..., \infty\}$, in the following manner:

$$|z> = \frac{1}{\sqrt{N(|z|^2)}} \sum_{n=0}^{\infty} \frac{z^n}{\sqrt{\rho(n)}} |n> \qquad (1.1)$$

where $\rho(n)$ are some functions depending on the main quantum number $n$, as well as on some constants. This is an extremely important entity, because it determines the internal structure of the CSs and the $\rho(n)$ entities were called the structure constants of CSs.

The structure constants $\rho(n)$ also determine the expression of the normalization function of CSs $N(|z|^2)$, which is obtained from the normalization condition for the CSs, $<z|z> = 1$. Let's point out what generally characterizes "non-canonical" CSs.

The nonlinear coherent states (NCSs) are states generated by the deformed annihilation operator $\hat{\mathcal{A}}_- = \hat{a}\sqrt{f(\hat{n})}$, where $f(\hat{n})$ is the deformation function depending on the particle number operator $\hat{n}$, with $\hat{n}|n> = n|n>$ and $f(\hat{n})|n> = f(n)|n>$. Consequently, the structure constants depend on the deformation function, $\rho(n; f)$. At the limit $f(\hat{n}) \to 1$, the NCSs turn into canonical CSs.

The q-deformed coherent states (qCSs) are states based on the modified or q-deformed commutation relation $\hat{a}\hat{a}^+ - q\hat{a}^+\hat{a} = 1$, where $q$ is a real parameter $0 < q < 1$. Their structure constants depend also on the deformation parameter, $\rho(n; q)$, and at the limit $q \to 1$ we obtain the CCSs.

The generalized hypergeometric coherent states (GHCSs) are states with a more pronounced generalization, generated by the nonlinear annihilation $\hat{\mathcal{A}}_-$ and creation $\hat{\mathcal{A}}_+$ operators. The normalization function of these CSs is a generalized hypergeometric function: $N(|z|^2) = {}_pF_q\left(\{a_i\}_1^p; \{b_j\}_1^q; |z|^2\right)$, where $\{a_i\}_1^p$ is a shorthand notation for the set of real parameters $\{a_i\}_1^p \equiv \{a_1, a_2, ..., a_p\}$. Obviously their structure constants are some functions of the form $\rho\left(n; \{a_i\}_1^p, \{b_j\}_1^q\right)$. At the limit $p = q$ and $\{a_i\}_1^p = \{b_j\}_1^q$ we recover the canonical CSs.

But in mathematics there are special functions of greater generalization than generalized hypergeometric functions, for example Fox-Wright (F-W) functions (also known as Fox–Wright Psi functions) [5], [6]. Several generating functions for some classes of functions associated to the Fox-Wright functions are studied in [7]. Apart from the many applications of Fox-Wright functions in mathematics [8], [9], [10], as well as in physics (anomalous transport in physics, fractional diffusion and sub-diffusion phenomena) [11], [12]) only recently has the issue of the connection between these functions and coherent states come into focus. In this connection, F-W functions play the role of normalization functions of coherent states [13], [14].

That is why the main purpose of this paper is to construct and study the properties of coherent states whose normalizing function is precisely the F-W function. We will refer to both pure coherent states and mixed (thermal) ones. With this paper we hope to broaden the applicability of the F-W functions, this time in quantum mechanics and quantum optics, where coherent states represent an important entity.

For these coherent states we will use the name "Fox-Wright coherent states (F-WCSs)" for the simple reason that the normalized function of these coherent states is the Fox-Wright function itself.

Moreover, the definition of the F-W function is

$$
{}_p\Psi_q\left[\begin{matrix}(a_1,\,A_1)\,(a_2,\,A_2)\,...\,(a_p,\,A_p)\\(b_1,\,B_1)\,(b_2,\,B_2)\,...\,(b_q,\,B_q)\end{matrix}\,;\;z\right]=\sum_{n=0}^{\infty}\frac{\prod_{i=1}^{p}\Gamma(a_i+A_i\,n)}{\prod_{j=1}^{q}\Gamma(b_j+B_j\,n)}\frac{z^n}{n!}
\tag{1.2}
$$

where the numbers $a_i+A_i\,n$ and $b_j+B_j\,n$ are complex or real, but non-integers.

The symbol $\Gamma(x+y\,n)$ has the following meaning:

$$
\Gamma(x+y\,n)=(x+y(n-1))(x+y(n-2))...(x+y(n-n))\Gamma(x)=
$$
$$
=y^n\left(\frac{x}{y}+n-1\right)\left(\frac{x}{y}+n-2\right)...\left(\frac{x}{y}+n-n\right)\Gamma(x)
\tag{1.3}
$$

$$
\Gamma(x+y\,n)=y^n\,\frac{\Gamma\left(\dfrac{x}{y}+n\right)}{\Gamma\left(\dfrac{x}{y}\right)}\,\Gamma(x)
\tag{1.4}
$$

In order to simplify the formulas, but also for reasons of practical applications in real models of the F-W functions, in the following we will consider that the parameters $a_i$, $A_i$, $b_j$, $B_j$ are positive numbers. As a result, Euler's Gamma functions, which intervene in the definition of the Fox-Wright functions, will also be positive, for all values of $n$.

In order to reduce the size of the formulas, we will continue to use the following notation, only where there is no danger of confusion

$$
{}_p\Psi_q\left[\begin{matrix}(a_1,\,A_1)\,(a_2,\,A_2)\,...\,(a_p,\,A_p)\\(b_1,\,B_1)\,(b_2,\,B_2)\,...\,(b_q,\,B_q)\end{matrix}\,;\;z\right]\equiv{}_p\Psi_q\left(\begin{matrix}(a,A)\\(b,B)\end{matrix}\bigg|\,z\right)
\tag{1.5}
$$

where $(a,A)\equiv\left\{(a_i,\,A_i)_1^p\right\}\equiv\left\{(a_1,\,A_1)\,(a_2,\,A_2)\,...\,(a_p,\,A_p)\right\}$ and so on.

The differences, in terms of the mathematical structure, of these types of CSs start from the different choice of the expression of the function $e\left(n\left|\begin{matrix}(a,A)\\(b,B)\end{matrix}\right.\right)\equiv e(n)$ that appears in the relation that represents the action of the annihilation operator on the Fock vector:

$$\hat{\mathcal{A}}_{-}\mid n>=\sqrt{e(n)}\mid n-1>$$ (1.6)

respectively from the definition of CSs (in the Barut-Girardello manner [15])

$$\hat{\mathcal{A}}_{-}\mid z>=z\mid z>$$ (1.7)

There is the following relationship between the structure constants $\rho(n)$ and the functions $e(n)$:

$$\rho(n)=\prod_{j=1}^{n}e(j)=n!\frac{\prod_{j=1}^{q}\Gamma\left(b_{j}+B_{j}\,n\right)}{\prod_{i=1}^{p}\Gamma\left(a_{i}+A_{i}\,n\right)}\equiv\rho\left(n\left|\begin{matrix}(a,A)\\(b,B)\end{matrix}\right.\right)$$ (1.8)

It is not difficult to observe the following relation between the structural constants and the F-W function with argument unity.

$$\sum_{n=0}^{\infty}\frac{1}{\rho\left(n\left|\begin{matrix}(a,A)\\(b,B)\end{matrix}\right.\right)}={}_{p}\Psi_{q}\left(\begin{matrix}(a,A)\\(b,B)\end{matrix}\right|\left.1\right)$$ (1.9)

## 2. Some properties of the Fox-Wright function

The F-W function is in fact a generalisation of the generalised hypergeometric function ${}_{p}F_{q}\left(\{a_{i}\}_{1}^{p};\{b_{j}\}_{1}^{q};z\right)$ and was initially introduced by Fox (in 1928) [5] and reintroduced by Wright in 1935 [6]. Compared to the hypergeometric function, the generalization consists in the fact that, instead of the sets of coefficients $\{a_{i}+n\}_{1}^{p}$ and $\{b_{j}+n\}_{1}^{q}$, the Fox-Wright function has the generalized sets $\{a_{i}+A_{i}\,n\}_{1}^{p}$ and $\{b_{j}+B_{j}\,n\}_{1}^{q}$.

With the notations introduced previously, the F-W function is

$$_{p}\Psi_{q}\left(\begin{matrix}(a,A)\\(b,B)\end{matrix}\right|\left.z\right)=\sum_{n=0}^{\infty}\frac{z^{n}}{{}_{p}\rho_{q}(n)}$$ (2.1)

as well as

$$_{p}\rho_{q}(n)\equiv n!\frac{\prod_{j=1}^{q}\Gamma\left(b_{j}+B_{j}\,n\right)}{\prod_{i=1}^{p}\Gamma\left(a_{i}+A_{i}\,n\right)}$$ (2.2)

Moreover, the F-W function can be defined also as an integral representation in complex $z$ space

$$_p\Psi_q\!\left(\!\left.\begin{matrix}(a,A)\\(b,B)\end{matrix}\right|z\right)=\frac{1}{2\pi i}\int_{\mathcal{L}}\frac{\prod_{j=1}^{m}\Gamma(b_j+B_j\,s)\prod_{i=1}^{n}\Gamma(1-a_i-A_i\,s)}{\prod_{j=m+1}^{q}\Gamma(1-b_j-B_j\,s)\prod_{i=n+1}^{p}\Gamma(a_i+A_i\,s)}\,z^{-s}ds \qquad (2.3)$$

where $1\le n\le p$ and $1\le m\le q$ are integers, as in the case of hypergeometric function. This function is defined for every $z\in C\setminus\{0\}$, $a_i$, $b_j\in C$ and $A_i$, $B_j\in\mathcal{R}$. As an observation, the F-W function, as well as the Mittag-Leffler function are the particular cases of the Fox $H$ function [13]

The above series is convergent if [14]

$$\sum_{j=1}^{q}B_j-\sum_{i=1}^{p}A_i\quad\begin{cases}>-1, & \text{convergent in the entire complex } z \text{ plane}\\ =-1, & \text{convergent for every bounded } /z/\end{cases} \qquad (2.4)$$

The $m$-th derivative of the F-W function is

$$\left(\frac{d}{dz}\right)^m{}_p\Psi_q\!\left(\!\left.\begin{matrix}(a,A)\\(b,B)\end{matrix}\right|z\right)=\sum_{n=0}^{\infty}\frac{\prod_{i=1}^{p}\Gamma(a_i+A_i\,m+A_i\,n)}{\prod_{j=1}^{q}\Gamma(b_j+B_j\,m+B_j\,n)}\frac{z^n}{n!}\equiv{}_p\Psi_q\!\left(\!\left.\begin{matrix}(a+A\cdot m,A)\\(b+B\cdot m,B)\end{matrix}\right|z\right) \qquad (2.5)$$

Particularly, for $z\to-z$, $p=q$ and $B_i\to A_i$, we will rediscover a corresponding formula from [9].

On the other hand, by particularizing the numbers $p$ and $q$, as well as the coefficients $a_i$, $A_i$, $b_j$, $B_j$ from the definition above, several special functions can be obtained (for example, the Wright function, Bessel function, hypergeometric function and Mittag-Leffler function).

In this paper we will only give the multi-index Mittag-Leffler function as an example, which allows, among other things, the formulation of corresponding Mittag-Leffler coherent states (M-LCSs) [16].

Starting from the initial definition of the Mittag-Leffler function [17], Srivastava and Tomovski defined the generalized Mittag-Leffler function, in the form [18]

$$E_{\beta,\alpha}^{\gamma,k}(z)=\sum_{n=0}^{\infty}\frac{(\gamma)_{n,k}}{\Gamma_\alpha(\beta+\alpha\,n)}\frac{z^n}{n!},\quad z,\alpha,\beta,\gamma\in C,\ \mathcal{R}e(\alpha)>0,\ \mathcal{R}e(k)>0 \qquad (2.6)$$

where here appear the generalized Pochhammer symbols (or Pochhammer $k$-symbols), having $n$ as discontinuous variable, e.g.[19]:

$$(x)_{n,y}=x(x+y)(x+2y)...(x+(n-1)y)=\frac{\Gamma(x+yn)}{\Gamma(x)}=y^n\left(\frac{x}{y}\right)_n=y^n\frac{\Gamma\!\left(\dfrac{x}{y}+n\right)}{\Gamma\!\left(\dfrac{x}{y}\right)} \qquad (2.7)$$

The next logical step in generalizing the Mittag-Leffler function, that is, to define this function to contain multiple indices. Starting from the definition of the function with 4 indices, [20]

$$E_{\beta,\alpha}^{\gamma,k}(z)=\frac{1}{\Gamma(\beta)}\sum_{n=0}^{\infty}\frac{(\gamma)_{n,k}}{(\beta)_{n,\alpha}}\frac{z^n}{n!} \qquad (2.8)$$

using the following substitutions: $\gamma \to \{a_i\}_1^p$, $k \to A_i$, $\beta \to \{b_j\}_1^q$ and $k \to B_j$, it is easily obtained the multi-index Mittag-Leffler function, defined as

$$E_{\{b_j\}_1^q, B_j}^{\{a_i\}_1^p, A_i}(z) = \sum_{n=0}^{\infty} \frac{\prod_{i=1}^{p} (a_i)_{n, A_i}}{\prod_{j=1}^{q} \Gamma(b_j + B_j n)} \frac{z^n}{n!} = \frac{1}{\prod_{i=1}^{p} \Gamma(a_i)} \sum_{n=0}^{\infty} \frac{\prod_{i=1}^{p} \Gamma(a_i + A_i n)}{\prod_{j=1}^{q} \Gamma(b_j + B_j n)} \frac{z^n}{n!} \qquad (2.9)$$

It is not difficult to see that this function is proportional to the F-W function:

$$_p\Psi_q\left(\begin{matrix}(a, A) \\ (b, B)\end{matrix} \middle| z\right) = \prod_{i=1}^{p} \Gamma(a_i) E_{\{b_j\}_1^q, B_j}^{\{a_i\}_1^p, A_i}(z) \qquad (2.10)$$

Since the coherent states associated with the multi-index Mittag-Leffler function [16] have been formulated, it is interesting to see what the coherent states associated with the F-W function look like and what properties they have. This is, in fact, one of the goals of this paper.

On the other hand, the F-W function is proportional with the hypergeometric function.

$$_p\Psi_q\left(\begin{matrix}(a, A) \\ (b, B)\end{matrix} \middle| z\right) = \sum_{n=0}^{\infty} \frac{\prod_{i=1}^{p} \Gamma(a_i + A_i n)}{\prod_{j=1}^{q} \Gamma(b_j + B_j n)} \frac{z^n}{n!} = \frac{\prod_{i=1}^{p} \Gamma(a_i)}{\prod_{j=1}^{q} \Gamma(b_j)} \sum_{n=0}^{\infty} \frac{\prod_{i=1}^{p} \left(\frac{a_i}{A_i}\right)_n}{\prod_{j=1}^{q} \left(\frac{b_j}{B_j}\right)_n} \frac{\left(\frac{\prod_{i=1}^{p} A_i}{\prod_{j=1}^{q} B_j} z\right)^n}{n!} \qquad (2.11)$$

so that, finally

$$_p\Psi_q\left(\begin{matrix}(a, A) \\ (b, B)\end{matrix} \middle| z\right) = \frac{\prod_{i=1}^{p} \Gamma(a_i)}{\prod_{j=1}^{q} \Gamma(b_j)} \, _pF_q\left(\left\{\frac{a_i}{A_i}\right\}_1^p \; ; \; \left\{\frac{b_j}{B_j}\right\}_1^q \; ; \; \frac{\prod_{i=1}^{p} A_i}{\prod_{j=1}^{q} B_j} z\right) \qquad (2.12)$$

From here it follows immediately a particular case, i.e. if $A_i$ and $B_j$ are equal to 1, the Fox-Wright function differs from the generalized hypergeometric function only by a constant multiplier [8]

$$_p\Psi_q\left(\begin{matrix}(a, 1) \\ (b, 1)\end{matrix} \middle| z\right) = \frac{\prod_{i=1}^{p} \Gamma(a_i)}{\prod_{j=1}^{q} \Gamma(b_j)} \, _pF_q\left(\{a_i\}_1^p \; ; \; \{b_j\}_1^q \; ; \; z\right) \qquad (2.13)$$

In the following we will use a F-W function $_p\Psi_q\left(\begin{matrix}(a, A) \\ (b, B)\end{matrix} \middle| |z|^2\right)$, where the variable $z$ is complex. Then, let's we evaluate the Laplace transform of this function will be

$$\mathcal{L}\left\{{}_p\Psi_q\right\}(s)=\int_0^\infty d\left(|z|^2\right)e^{-s|z|^2}\ {}_p\Psi_q\left(\begin{matrix}(a,A)\\(b,B)\end{matrix}\middle|\ |z|^2\right)=\sum_{n=0}^\infty\frac{\prod\limits_{i=1}^p\Gamma\left(a_i+A_i\,n\right)}{\prod\limits_{j=1}^q\Gamma\left(b_j+B_j\,n\right)}\frac{1}{n!}\int_0^\infty d\left(|z|^2\right)e^{-s|z|^2}\left(|z|^2\right)^n=$$

$$=\frac{1}{s}\sum_{n=0}^\infty\frac{\Gamma\left(1+n\right)\prod\limits_{i=1}^p\Gamma\left(a_i+A_i\,n\right)}{\prod\limits_{j=1}^q\Gamma\left(b_j+B_j\,n\right)}\frac{\left(s^{-1}\right)^n}{n!}=\frac{1}{s}\ {}_{p+1}\Psi_q\left(\begin{matrix}(a,A),(1,1)\\(b,B)\end{matrix}\middle|\ \frac{1}{s}\right)$$

<div align="right">(2.14)</div>

This is a particular case of the result obtained in [12]. The Laplace transform is used to obtain the solutions of a generalized diffusion equation, called the time-fractional diffusion equation of order $\alpha$ (where $0<\alpha\le 1$) [11]:

$$\frac{\partial^\alpha}{\partial t^\alpha}u(x,t)=\frac{\partial^2}{\partial t^2}u(x,t),\ \ u(x,0)=\delta(x),\ \ \ x\in\mathcal{R},\ t\ge 0, \tag{2.15}$$

## 3. Diagonal operators ordering technique and the Fox-Wright coherent states

In quantum mechanics and, in particular, in quantum optics, three ways of arranging quantum operators are used, when products of operators are involved: normal ordering, antinormal ordering, and Weyl or symmetric ordering. A new technique for calculating normally ordered forms of unitary operators which induce symplectic transformations was introduced by Fan and called Integration Within an Ordered Product (IWOP) of operators, with the use of Dirac's bra-ket formalism (see, e.g. [21] and the references therein). Until now, the IWOP technique has been used in a relatively limited area of applicability, in the sense that it involves the use of only the anihillation $\hat{a}$ and creation $\hat{a}_+$ canonical operators. These operators are characteristics of the one-dimensional linear harmonic oscillator (HO-1D) and for this reason these operators are considered linear operators.

In order to apply this integration technique also to nonlinear quantum systems, we generalized Fan's procedure using a pair of creation $\hat{\mathcal{A}}_+$ and annihilation $\hat{\mathcal{A}}_-$ operators, that act on the Fock vectors $|n>$ in the space attached to the examined quantum system. The calculation rules are identical to those in the IWOP case, but for the sake of distinction, we have named them the diagonal operators ordering technique (DOOT), making extensive use of the properties of generalized hypergeometric functions and using the symbol # # [22]. Consequently, IWOP appears as a particular case of DOOT. The rules of a normal ordering technique - diagonal operators ordering technique (DOOT), consist of the following:

*a)* Inside the symbol # # the order of the operators $\hat{\mathcal{A}}_-$ and $\hat{\mathcal{A}}_+$ can be permuted like commutable operators, so that finally will result a normal ordering of operators: $\hat{\mathcal{A}}_+$ on the left, and $\hat{\mathcal{A}}_-$ on the right;

*b)* A symbol # # inside another symbol # # can be deleted;

*c)* A normally ordered product of operators can be integrated or differentiated, with respect to *c*-numbers, according to the usual rules. In addition, the operators $\hat{\mathcal{A}}_-$ and $\hat{\mathcal{A}}_+$ are considered as being simple *c*-numbers and can be taken out from the symbol # #;

*d)* The projector $|0><0|$ of the normalized vacuum state $|0>$, in the frame of DOOT, is the reciprocal function $\left(1/N\left(\hat{\mathcal{A}}_+\hat{\mathcal{A}}_-\right)\right)$ of the normalization function of CSs $N\left(\hat{\mathcal{A}}_+\hat{\mathcal{A}}_-\right)=_pF_q\left(\{a_i\}_1^p\,;\,\{b_j\}_1^q\,;\hat{\mathcal{A}}_+\hat{\mathcal{A}}_-\right)$, in normal order, i. e. the reciprocal function of normal ordered generalized hypergeometric function which has as argument the normal product of operators $\hat{\mathcal{A}}_+\hat{\mathcal{A}}_-$:

$$|0><0|=\#\frac{1}{_pF_q\left(\{a_i\}_1^p\,;\,\{b_j\}_1^q\,;\hat{\mathcal{A}}_+\hat{\mathcal{A}}_-\right)}\#\qquad(3.1)$$

By particularizing the integers $p$ and $q$, as well as the sets of real numbers $\{a_i\}_1^p$ and $\{b_j\}_1^q$, the generalized hypergeometric function becomes equal to other functions, characteristic of the examined system and the related CSs.

Now, let's consider that a pair of operators anihillation $\hat{\mathcal{A}}_-$ and creation $\hat{\mathcal{A}}_+$ are non-linear, defined as

$$\hat{\mathcal{A}}_-=\hat{a}\sqrt{_pf_q(\hat{n})}\quad,\;\hat{\mathcal{A}}_+=\sqrt{_pf_q(\hat{n})}\,\hat{a}^+=\hat{a}^+\sqrt{_pf_q(\hat{n}+1)}\qquad(3.2)$$

The non-linearity or deformation function $_pf_q(\hat{n})$ are dependent on the particle number operator $\hat{n}$, and of the sets of real numbers $\left\{(a_i,\,A_i)_1^p\right\}\equiv(\text{a},\text{A})$ and $\left\{(b_j,\,B_j)_1^q\right\}\equiv(\text{b},\text{B})$, i.e. $_pf_q(\hat{n})=_pf_q\left(\hat{n},\,(\text{a},\text{A}),(\text{b},\text{B})\right)$, but for the brevity we will write only $_pf_q(\hat{n})$. Its eigenvalue equation is

$$_pf_q(\hat{n})|\,n>=_pf_q(n)|\,n>\qquad(3.3)$$

These operators act on the vector's orthonormal basis $\{|\,n>,\;n=0,1,2,...,\infty\}$, in the following way

$$\hat{\mathcal{A}}_-|\,n>=\sqrt{_pe_q(n)}\,|\,n-1>,\;\;<n|\,\hat{\mathcal{A}}_+=\sqrt{_pe_q(n+1)}<n+1|\;\;,\;\;\hat{\mathcal{A}}_+\hat{\mathcal{A}}_-|\,n>=_pe_q(n)|\,n>\quad(3.4)$$

$$<0|\left(\hat{\mathcal{A}}_+\right)^n=\sqrt{\prod_{j=1}^n{}_pe_q(j)}<n|\equiv\sqrt{_pg_q(n)}<n|\qquad(3.5)$$

If we choose the eigenvalues of the deformation function in the form

$$_pf_q(n)=\frac{\prod_{j=1}^q\left(b_j+B_j\cdot(n-1)\right)}{\prod_{i=1}^p\left(a_i+A_i\cdot(n-1)\right)}\qquad(3.6)$$

then the expression $_pe_q(n)$ of the eigenvalues of the product $\hat{\mathcal{A}}_+\hat{\mathcal{A}}_-$ becomes

$$_p e_q(n) = n_p f_q(n) = n \frac{\prod\limits_{j=1}^{q}\left(b_j + B_j \cdot (n-1)\right)}{\prod\limits_{i=1}^{p}\left(a_i + A_i \cdot (n-1)\right)} \tag{3.7}$$

Obviously, different quantum systems have different values of the sets of real numbers $\left\{\left(a_i, A_i\right)_1^p\right\} \equiv (a, A)$ and $\left\{\left(b_j, B_j\right)_1^q\right\} \equiv (b, B)$.

Next, we will define coherent states in the sense of Barut and Girardello, according to Eq. (1.7), that is as eigenvectors of the annihilation operator $\hat{\mathcal{A}}$ [15].

Any coherent state can be expanded as a power series with respect to a set of orthogonal vectors. If we consider the basis of Fock vectors, we will have

$$| z > = \sum_{n=0}^{\infty} c_n(z) | n > \tag{3.8}$$

To find the expansion coefficients $c_n(z)$, we will use both the definition of coherent states and also the action of the annihilation operator on Fock vectors. So, we have

$$c_n(z) = c_0(z) \frac{z^n}{\sqrt{\prod\limits_{m=1}^{n} {}_p e_q(m)}} = c_0(z) \frac{z^n}{\sqrt{n! \prod\limits_{m=1}^{n} \frac{\prod\limits_{j=1}^{q}\left(b_j + B_j \cdot (m-1)\right)}{\prod\limits_{i=1}^{p}\left(a_i + A_i \cdot (m-1)\right)}}} \tag{3.9}$$

We now introduce the structure constants for CSs in the form

$$_p g_q(n) \equiv \prod_{m=1}^{n} {}_p e_q(m) = n! \prod_{m=1}^{n} \frac{\prod\limits_{j=1}^{q}\left(b_j + B_j \cdot (m-1)\right)}{\prod\limits_{i=1}^{p}\left(a_i + A_i \cdot (m-1)\right)} \tag{3.10}$$

and rewrite the numerator and the denominator of the last fraction so that the generalized Pochhammer symbols appear

$$\prod_{j=1}^{q}\prod_{m=1}^{n}\left(b_j + B_j \cdot (m-1)\right) = \prod_{j=1}^{q}\left(b_j + B_j \cdot 0\right)\left(b_j + B_j \cdot 1\right)\left(b_j + B_j \cdot 2\right)...\left(b_j + B_j \cdot (n-1)\right) =$$
$$= \prod_{j=1}^{q} \frac{\Gamma\left(b_j + B_j \cdot n\right)}{\Gamma\left(b_j\right)} = \prod_{j=1}^{q}\left(b_j\right)_{n, B_j} \tag{3.11}$$

With the above notations, the structure constants becomes, using Eq. (2.2)

$$_p g_q(n) = n! \frac{\prod\limits_{j=1}^{q}\left(b_j\right)_{n, B_j}}{\prod\limits_{i=1}^{p}\left(a_i\right)_{n, A_i}} = \frac{\prod\limits_{i=1}^{p}\Gamma\left(a_i\right)}{\prod\limits_{j=1}^{q}\Gamma\left(b_j\right)} n! \frac{\prod\limits_{j=1}^{q}\Gamma\left(b_j + B_j n\right)}{\prod\limits_{i=1}^{p}\Gamma\left(a_i + A_i n\right)} = \frac{\prod\limits_{i=1}^{p}\Gamma\left(a_i\right)}{\prod\limits_{j=1}^{q}\Gamma\left(b_j\right)} {}_p\rho_q(n) \tag{3.12}$$

Then the expression for CSs is

$$|z> = c_0(z) \sum_{n=0}^{\infty} \frac{z^n}{\sqrt{_p g_q(n)}} |n> \qquad (3.13)$$

The constant $c_0(z)$ is related to the normalization function which is determined from the normalization condition of coherent states $<z|z>=1$. Its square must be a positive definite and finite continuous function $0<|c_0(z)|^2<+\infty$, for every value of variable $z$:

$$\left[c_0(z)\right]^{-2} = \sum_{n=0}^{\infty} \frac{\left(|z|^2\right)^n}{_p g_q(n)} = \sum_{n=0}^{\infty} \frac{\prod_{i=1}^{p}(a_i)_{n,A_i}}{\prod_{j=1}^{q}(b_j)_{n,B_j}} \frac{\left(|z|^2\right)^n}{n!} = \frac{\prod_{j=1}^{q}\Gamma(b_j)}{\prod_{i=1}^{p}\Gamma(a_i)} \sum_{n=0}^{\infty} \frac{\prod_{i=1}^{p}\Gamma(a_i+A_i\,n)}{\prod_{j=1}^{q}\Gamma(b_j+B_j\,n)} \frac{\left(|z|^2\right)^n}{n!} =$$

$$= \frac{\prod_{j=1}^{q}\Gamma(b_j)}{\prod_{i=1}^{p}\Gamma(a_i)} {}_p\Psi_q\left(\begin{matrix}(a,A)\\(b,B)\end{matrix}\middle|\,|z|^2\right) \qquad (3.14)$$

Substituting in the above expression for CSs we will obtain the final expression for CSs, let's call them Fox-Wright coherent states (F-WCSs):

$$|z> = \frac{1}{\sqrt{_p\Psi_q\left(\begin{matrix}(a,A)\\(b,B)\end{matrix}\middle|\,|z|^2\right)}} \sum_{n=0}^{\infty} \sqrt{\frac{\prod_{i=1}^{p}\Gamma(a_i+A_i\,n)}{\prod_{j=1}^{q}\Gamma(b_j+B_j\,n)}} \frac{z^n}{\sqrt{n!}} |n> =$$

$$= \frac{1}{\sqrt{_p\Psi_q\left(\begin{matrix}(a,A)\\(b,B)\end{matrix}\middle|\,|z|^2\right)}} \sum_{n=0}^{\infty} \frac{z^n}{\sqrt{_p\rho_q(n)}} |n> \qquad (3.15)$$

It is observed that the normalization function of the F-WCSs is just the F-W function, i.e.

$$N\left(|z|^2\right) = {}_p\Psi_q\left(\begin{matrix}(a,A)\\(b,B)\end{matrix}\middle|\,|z|^2\right) \qquad (3.16)$$

In this case, their structure constants are some functions of the form $\rho\big(n;(a,A),(b,B)\big) \equiv {}_p\rho_q(n)$. At the limit $\left\{(a_i,A_i)_1^p\right\} = \left\{(b_j,B_j)_1^q\right\}$ will be obtained the canonical coherent states (CCSs).

Referring to Eqs. (3.5) and (3.12), by substitution we get

$$| z> = \sqrt{\frac{\prod_{j=1}^{q} \Gamma(b_j)}{\prod_{i=1}^{p} \Gamma(a_i)}} \frac{1}{\sqrt{{}_p \Psi_q \left( \begin{matrix} (a,A) \\ (b,B) \end{matrix} \middle| \, | z |^2 \right)}} \sum_{n=0}^{\infty} \frac{\left( z \hat{\mathcal{A}}_+ \right)^n}{{}_p \rho_q(n)} | 0 > =$$

$$= \sqrt{\frac{\prod_{j=1}^{q} \Gamma(b_j)}{\prod_{i=1}^{p} \Gamma(a_i)}} \frac{1}{\sqrt{{}_p \Psi_q \left( \begin{matrix} (a,A) \\ (b,B) \end{matrix} \middle| \, | z |^2 \right)}} \, {}_p \Psi_q \left( \begin{matrix} (a,A) \\ (b,B) \end{matrix} \middle| \, z \hat{\mathcal{A}}_+ \right) | 0 > \tag{3.17}$$

Their counterpart bra vector (in Dirac's language) $< z |$ is

$$< z | = \sqrt{\frac{\prod_{j=1}^{q} \Gamma(b_j)}{\prod_{i=1}^{p} \Gamma(a_i)}} \frac{1}{\sqrt{{}_p \Psi_q \left( \begin{matrix} (a,A) \\ (b,B) \end{matrix} \middle| \, | z |^2 \right)}} < 0 |_p \Psi_q \left( \begin{matrix} (a,A) \\ (b,B) \end{matrix} \middle| \, z^* \hat{\mathcal{A}}_- \right) \tag{3.18}$$

so that the projector on the coherent state $| z > < z |$ will be

$$| z > < z | = \frac{\prod_{j=1}^{q} \Gamma(b_j)}{\prod_{i=1}^{p} \Gamma(a_i)} \frac{\# \, {}_p \Psi_q \left( \begin{matrix} (a,A) \\ (b,B) \end{matrix} \middle| \, z \hat{\mathcal{A}}_+ \right) | 0 > < 0 | \, {}_p \Psi_q \left( \begin{matrix} (a,A) \\ (b,B) \end{matrix} \middle| \, z^* \hat{\mathcal{A}}_- \right) \#}{{}_p \Psi_q \left( \begin{matrix} (a,A) \\ (b,B) \end{matrix} \middle| \, | z |^2 \right)} \tag{3.19}$$

Using the DOOT rules, let's find the expression for the vacuum projector $| 0 > < 0 |$. For this, let's start from the closing relation of the Fock vectors $\sum_{n=0}^{\infty} | n > < n | = 1$.

We will get, successively

$$\sum_{n=0}^{\infty} | n > < n | = \# \sum_{n=0}^{\infty} \frac{\left( \hat{\mathcal{A}}_+ \right)^n}{\sqrt{{}_p g_q(n)}} | 0 > < 0 | \frac{\left( \hat{\mathcal{A}}_- \right)^n}{\sqrt{{}_p g_q(n)}} \# =$$

$$= | 0 > < 0 | \, \# \sum_{n=0}^{\infty} \frac{\left( \hat{\mathcal{A}}_+ \hat{\mathcal{A}}_- \right)^n}{{}_p g_q(n)} \# = | 0 > < 0 | \frac{\prod_{j=1}^{q} \Gamma(b_j)}{\prod_{i=1}^{p} \Gamma(a_i)} \# \, {}_p \Psi_q \left( \begin{matrix} (a,A) \\ (b,B) \end{matrix} \middle| \, \hat{\mathcal{A}}_+ \hat{\mathcal{A}}_- \right) \# = 1 \tag{3.20}$$

Consequently, the final expression for the vacuum projector becomes

$$| 0 > < 0 | = \frac{\prod_{i=1}^{p} \Gamma(a_i)}{\prod_{j=1}^{q} \Gamma(b_j)} \# \frac{1}{{}_p \Psi_q \left( \begin{matrix} (a,A) \\ (b,B) \end{matrix} \middle| \, \hat{\mathcal{A}}_+ \hat{\mathcal{A}}_- \right)} \# \tag{3.21}$$

This result is in agreement, on the one hand, by considering $A = B = 1$ with those obtained previously, Eq. (3.1) as general case [22], and for the other hand for the particular case of Mittag-Leffler coherent states [16].

$$| 0 > < 0 | = \frac{\prod_{i=1}^{p} \Gamma(a_i)}{\prod_{j=1}^{q} \Gamma(b_j)} \# \frac{1}{{}_{p}\Psi_{q}\left(\begin{matrix}(a,A)\\(b,B)\end{matrix}\middle| \hat{\mathcal{A}}_{+}\hat{\mathcal{A}}_{-}\right)} \# =$$

$$= \frac{1}{\prod_{j=1}^{q} \Gamma(b_j)} \# \frac{1}{E_{(b_j)_1^q,B_j}^{\{a_i\}_1^p,A_i}\left(\hat{\mathcal{A}}_{+}\hat{\mathcal{A}}_{-}\right)} \# = \# \frac{1}{{}_{p}F_{q}\left(\left\{\frac{a_i}{A_i}\right\}_1^p ; \left\{\frac{b_j}{B_j}\right\}_1^q ; \hat{\mathcal{A}}_{+}\hat{\mathcal{A}}_{-}\right)} \#$$

(3.22)

With this result the projector on the state $| z >$ is written as

$$| z > < z | = \frac{1}{{}_{p}\Psi_{q}\left(\begin{matrix}(a,A)\\(b,B)\end{matrix}\middle| |z|^2\right)} \# \frac{{}_{p}\Psi_{q}\left(\begin{matrix}(a,A)\\(b,B)\end{matrix}\middle| z\hat{\mathcal{A}}_{+}\right) {}_{p}\Psi_{q}\left(\begin{matrix}(a,A)\\(b,B)\end{matrix}\middle| z^*\hat{\mathcal{A}}_{-}\right)}{{}_{p}\Psi_{q}\left(\begin{matrix}(a,A)\\(b,B)\end{matrix}\middle| \hat{\mathcal{A}}_{+}\hat{\mathcal{A}}_{-}\right)} \#$$

(3.23)

The overlap of two F-WCSs is

$$< z | z' > = \frac{{}_{p}\Psi_{q}\left(\begin{matrix}(a,A)\\(b,B)\end{matrix}\middle| z^* z'\right)}{\sqrt{{}_{p}\Psi_{q}\left(\begin{matrix}(a,A)\\(b,B)\end{matrix}\middle| |z|^2\right)} \sqrt{{}_{p}\Psi_{q}\left(\begin{matrix}(a,A)\\(b,B)\end{matrix}\middle| |z'|^2\right)}}$$

(3.24)

Among papers similar to this problem, let us highlight the study of the generalization and properties of coherent states involving the Fox H function, as a more general case of the F-W function [13]. The name "coherent states of the Fox-Wright type (F-WCSs)", which we adopted, is justified by the fact that the normalization function of these CSs is precisely the F-W function.

## 4. Properties of the Fox-Wright coherent states

Let's check whether F-WCSs meet all the requirements imposed on CSs (the so-called "Klauder's minimal prescriptions") [23]:

**a) *Any set of CSs must be continuous in the complex label $z$.***

The vectors $| z >$ are the strong continuous functions of label $z \neq 0$, i.e. for every convergent labels $z'$, $z \in C$ such that $z' \to z$ it follows that $\| |z'> - |z> \| \to 0$. Continuity is also validated due to the implicit property that the series involved in the definition of CSs is a continuous function of the complex variable $z$ for any $z \in C$.

**b) *Any set of the CSs must be normalizable but non-orthogonal:***

$$\lim_{z' \to z} < z | z' > = \begin{cases} 1 , & \text{normalization condition} \\ \neq 0 & \text{non} - \text{orthogonality condition} \end{cases}$$

(4.1)

This property immediately follows from examining the expression of the overlap of two CSs, indicated above.

**c) *Any set of CSs must ensure the decomposition of the unit operator* or, in other words, *the any class of CSs must resolve the identity operator over the Fock vectors's basis.***
The problem here is to find a positive and continuous weight function $h\left(|z|^2\right)$ of the integration measure $d\mu(z) = \dfrac{d^2 z}{\pi} h\left(|z|^2\right) = \dfrac{d\varphi}{2\pi} d\left(|z|^2\right) h\left(|z|^2\right)$ in $\mathcal{R}$, which is specific for each type of CSs.

$$\int d\mu(z)\,|z><z| = I \tag{4.2}$$

After appropriate substitutions, we will have

$$\sum_{n,m=0}^{\infty} \frac{|n><m|}{\sqrt{_p\rho_q(n)}\sqrt{_p\rho_q(m)}} \int_0^\infty d\left(|z|^2\right) \frac{h\left(|z|^2\right)}{_p\Psi_q\left(\begin{matrix}(a,A)\\(b,B)\end{matrix}\Big||z|^2\right)} \int_0^{2\pi} \frac{d\varphi}{2\pi}\left(z^*\right)^n z^m = 1 \tag{4.3}$$

Since the angular integral has the value $\left(|z|^2\right)^n \delta_{nm}$, to satisfy the completeness relation of the Fock's vectors, $\sum_{n=0}^{\infty}|n><n| = 1$, we will need to solve the following integral in $\mathcal{R}$:

$$\int_0^\infty d\left(|z|^2\right) \frac{h\left(|z|^2\right)}{_p\Psi_q\left(\begin{matrix}(a,A)\\(b,B)\end{matrix}\Big||z|^2\right)}\left(|z|^2\right)^n = {}_p\rho_q(n) = n!\,\frac{\prod_{j=1}^{q}\Gamma\left(b_j + B_j\,n\right)}{\prod_{i=1}^{p}\Gamma\left(a_i + A_i\,n\right)} \tag{4.4}$$

After the transformation (see, Eq. (3.3)) as well as the substitution $n = s-1$, and with the observation that the weight function must be proportional to the Fox-Weight function $h\left(|z|^2\right) = \tilde{h}\left(|z|^2\right){}_p\Psi_q\left(\ldots\,;\,|z|^2\right)$, we arrive at the classical moment problem which implies as a solution a Meijer $G$-function [24]:

$$\int_0^\infty d\left(|z|^2\right)\tilde{h}\left(|z|^2\right)\left(|z|^2\right)^{s-1} = \frac{\prod_{i=1}^{p}\Gamma\left(\dfrac{a_i}{A_i}\right)}{\prod_{i=1}^{p}\Gamma(a_i)}\frac{\prod_{j=1}^{q}\Gamma(b_j)}{\prod_{j=1}^{q}\Gamma\left(\dfrac{b_j}{B_j}\right)}\left(\frac{\prod_{j=1}^{q}B_j}{\prod_{i=1}^{p}A_i}\right)^{s-1}\Gamma(s)\frac{\prod_{j=1}^{q}\Gamma\left(\dfrac{b_j}{B_j}-1+s\right)}{\prod_{i=1}^{p}\Gamma\left(\dfrac{a_i}{A_i}-1+s\right)} \tag{4.5}$$

It is useful to introduce the following notations for constant that appears above:

$$C_{p,q}(a,A,b,B) \equiv \frac{\prod_{i=1}^{p}A_i}{\prod_{j=1}^{q}B_j}\frac{\prod_{i=1}^{p}\Gamma\left(\dfrac{a_i}{A_i}\right)}{\prod_{i=1}^{p}\Gamma(a_i)}\frac{\prod_{j=1}^{q}\Gamma(b_j)}{\prod_{j=1}^{q}\Gamma\left(\dfrac{b_j}{B_j}\right)} \tag{4.6}$$

According to the Mellin inversion theorem of the Meijer's G function [24]

$$\int_0^\infty dx\, x^{s-1}\, G_{p,q}^{m,n}\!\left(\omega x \left| \begin{array}{ccc} \{a_i\}_1^n & ; & \{a_i\}_{n+1}^p \\ \{b_j\}_1^m & ; & \{b_j\}_{m+1}^q \end{array} \right.\right) = \frac{1}{\omega^s}\, \frac{\prod_{j=1}^m \Gamma(b_j+s)\prod_{i=1}^n \Gamma(1-a_i-s)}{\prod_{j=m+1}^q \Gamma(1-b_j-s)\prod_{i=n+1}^p \Gamma(a_i+s)} \qquad (4.7)$$

the solution is a function proportional with the Meijer $G$-function:

$$h\!\left(|z|^2\right) = C_{p,q}(a,A,b,B)\, {}_p\Psi_q\!\left(\ldots;|z|^2\right) G_{p,q+1}^{q+1,0}\!\left(\frac{\prod_{i=1}^p A_i}{\prod_{j=1}^q B_j}\Bigg| |z|^2 \left| \begin{array}{ccc} / & ; & \left\{\dfrac{a_i}{A_i}-1\right\}_1^p \\ 0\,,\,\left\{\dfrac{b_j}{B_j}-1\right\}_1^q & ; & / \end{array}\right.\right) \qquad (4.8)$$

Consequently, the integration measure becomes

$$d\mu(z) = C_{p,q}(a,A,b,B)\frac{d^2z}{\pi}\, {}_p\Psi_q\!\left(\left.\begin{array}{c}(a,A)\\(b,B)\end{array}\right| |z|^2\right) G_{p,q+1}^{q+1,0}\!\left(\frac{\prod_{i=1}^p A_i}{\prod_{j=1}^q B_j}\Bigg| |z|^2 \left| \begin{array}{ccc} / & ; & \left\{\dfrac{a_i}{A_i}-1\right\}_1^p \\ 0\,,\,\left\{\dfrac{b_j}{B_j}-1\right\}_1^q & ; & / \end{array}\right.\right)$$

$$(4.9)$$

Substituting this result into Eq. (5.3b), an important integral will result, which we will use in the following sections:

$$\int_0^\infty d\!\left(|z|^2\right) G_{p,q+1}^{q+1,0}\!\left(\frac{\prod_{i=1}^p A_i}{\prod_{j=1}^q B_j}\Bigg| |z|^2 \left| \begin{array}{ccc} / & ; & \left\{\dfrac{a_i}{A_i}-1\right\}_1^p \\ 0\,,\,\left\{\dfrac{b_j}{B_j}-1\right\}_1^q & ; & / \end{array}\right.\right)\left(|z|^2\right)^n =$$

$$(4.10)$$

$$= \frac{1}{C_{p,q}(a,A,b,B)}\, {}_p\rho_q(n) = \left(\frac{\prod_{j=1}^q B_j}{\prod_{i=1}^p A_i}\right)^n n!\,\frac{\prod_{j=1}^q \Gamma\!\left(\dfrac{b_j}{B_j}+n\right)}{\prod_{i=1}^p \Gamma\!\left(\dfrac{a_i}{A_i}+n\right)}$$

**Observation**. In the case $A_i = B_j = 1$ we recover the corresponding weight function of the generalized coherent states (GCSs) [22], respectively, for $a_i = b_j = 1$ and $A_i = B_j = 1$, we obtain $G_{0,1}^{1,0}\!\left(|z|^2|0\right) = \exp\!\left(-|z|^2\right)$ which correspond to the CCSs.

Let's multiply the decomposition relation of the unit operator on the left with $<z'|$ and on the right with $|z">$. We will have

$$\int d\mu(z)<z'|z><z|z">=<z'|z"> \qquad (4.11)$$

Using Eq. (5.1), we obtain a new relationship between the Fox-Wright functions:

$$_p\Psi_q\left(\left.\begin{matrix}(a,A)\\(b,B)\end{matrix}\right|z'^*\,z''\right)=C_{p,q}(a,A,b,B)\int\frac{d^2z}{\pi}\,G_{p,q+1}^{q+1,0}\left(\left.\begin{matrix}\prod\limits_{i=1}^p A_i\\\prod\limits_{j=1}^q B_j\end{matrix}\,|\,z\,|^2\right|\begin{matrix}\quad/\ ;&\left\{\dfrac{a_i}{A_i}-1\right\}_1^p\\0\,,\ \left\{\dfrac{b_j}{B_j}-1\right\}_1^q\ ;&/\end{matrix}\right)\times$$

$$\times\,_p\Psi_q\left(\left.\begin{matrix}(a,A)\\(b,B)\end{matrix}\right|z'^*\,z\right)\,_p\Psi_q\left(\left.\begin{matrix}(a,A)\\(b,B)\end{matrix}\right|z^*\,z''\right)$$

$$(4.12)$$

**d.) Any set of CSs** (associated with the quantum systems with infinite energy spectra) *satisfies the so-called "action identity" relation. The expected value* (**called the "lower symbol") of the Hamitonian mimics the classical energy-action relation** [4].

Generally, an expected value for a certain operator $\hat{O}$ is calculated as

$$<z|\,\hat{O}\,/z>=\frac{1}{_p\Psi_q\left(\left.\begin{matrix}(a,A)\\(b,B)\end{matrix}\right|\,|\,z\,|^2\right)}\sum_{n,m=0}^{\infty}\frac{\left(z^*\right)^n}{\sqrt{_p\rho_q(n)}}\frac{z^m}{\sqrt{_p\rho_q(m)}}<n|\,\hat{O}\,/m>\qquad(4.13)$$

Let us consider that the Hamiltonian of the system is equal to the normal product of the creation and annihilation operators $\hat{\mathcal{H}}=\hbar\omega\#\hat{\mathcal{A}}_+\hat{\mathcal{A}}_-\#$. Then, the dimensionless eigenvalue equation is

$$\#\hat{\mathcal{A}}_+\hat{\mathcal{A}}_-\#|n>=\,_pe_q(n)|\,n>\qquad(4.14)$$

and the expected value of the Hamiltonian in the F-WCSs representation has the expression

$$<z|\,\hat{\mathcal{H}}/z>=\hbar\omega\frac{1}{_p\Psi_q\left(\left.\begin{matrix}(a,A)\\(b,B)\end{matrix}\right|\,|\,z\,|^2\right)}\sum_{n=0}^{\infty}\frac{\left(|\,z\,|^2\right)^n}{_p\rho_q(n)}\,_pe_q(n)=\hbar\omega\frac{|\,z\,|^2}{_p\Psi_q\left(\left.\begin{matrix}(a,A)\\(b,B)\end{matrix}\right|\,|\,z\,|^2\right)}\sum_{n=0}^{\infty}\frac{\left(|\,z\,|^2\right)^{n-1}}{_p\rho_q(n-1)}\quad(4.15)$$

where we used the equality

$$\frac{1}{_p\rho_q(n)}\,_pe_q(n)=\frac{1}{_p\rho_q(n-1)}\qquad(4.16)$$

If we make the substitution $n-1=m$ and give up the non-physical field $m=-1$, sum is actually the Fox-Wright function and we get the final relation:

$$<z|\,\hat{\mathcal{H}}/z>=\hbar\omega\,|\,z\,|^2\qquad(4.17)$$

This represents the mathematical expression of the "action identity" (considering $\hbar=1$): the parameter $|\,z\,|^2$ can be interpret as the classical action variable conjugate with the angle variable $\omega$.

As a consequence of this result, the following expected value is obtained ($m$ is a positive integer):

$$<z|\#\left(\hat{\mathcal{A}}_+\hat{\mathcal{A}}_-\right)^m\#/z>=\left(|\,z\,|^2\right)^m\qquad(4.18)$$

that is, in calculating the average values in the F-WCSs representation, the result is that each product $\hat{\mathcal{A}}_+\hat{\mathcal{A}}_-$ is replaced by the variable $|\,z\,|^2$, i.e. $\hat{\mathcal{A}}_+\hat{\mathcal{A}}_-\rightarrow|\,z\,|^2$. This result also extends to a function that depends on:

$$< z \,|\# \, \mathcal{F}\left(\hat{\boldsymbol{\mathcal{A}}}_+ \hat{\boldsymbol{\mathcal{A}}}_-\right)\# |\, z> = \mathcal{F}\left(|\, z \,|^2\right) \tag{4.19}$$

*Observation:* The action identity it is valid only for coherent states of quantum systems that have a linear energy spectrum, therefore an infinite number of energy levels.

If we consider the limit

$$\lim_{f(\hat{\boldsymbol{n}}) \to 1} \# \, \hat{\boldsymbol{\mathcal{A}}}_+ \hat{\boldsymbol{\mathcal{A}}}_- \# = \hat{\boldsymbol{a}}^+ \hat{\boldsymbol{a}} = \hat{\boldsymbol{n}} \tag{4.20}$$

we can proceed to the calculation of Mandel's parameter which measures the deviation of the distribution of the average values of the particle number operator from the Poissonian statistics. The canonical CSs, i.e. those associated with the one-dimensional harmonic oscillator follow the Poisson distribution or Poisson statistics. This distribution is characterized by the variance of the number operator being equal to its average.

Let's evaluate Mandel's Q parameter by calculating the averages in the Fox-Wright coherent state representation. Their definition is [25]

$$Q_{|z|} \equiv \frac{< z |\, \hat{\boldsymbol{n}}^2 /z> - \left(< z |\, \hat{\boldsymbol{n}} /z>\right)^2}{< z |\, \hat{\boldsymbol{n}} /z>} - 1 \tag{4.21}$$

The expectation value of an integer power $s$ of number operator is

$$< z |\, \hat{\boldsymbol{n}}^s /z> = \frac{1}{{}_p\Psi_q\left(\begin{matrix}(\mathrm{a,A})\\(\mathrm{b,B})\end{matrix}\bigg|\,|\,z\,|^2\right)} \sum_{n=0}^{\infty} \frac{\left(|\,z\,|^2\right)^n}{{}_p\rho_q(n)} n^s =$$

$$= \frac{1}{{}_p\Psi_q\left(\begin{matrix}(\mathrm{a,A})\\(\mathrm{b,B})\end{matrix}\bigg|\,|\,z\,|^2\right)} \left(|\,z\,|^2 \frac{\partial}{\partial\,|\,z\,|^2}\right)^s {}_p\Psi_q\left(\begin{matrix}(\mathrm{a,A})\\(\mathrm{b,B})\end{matrix}\bigg|\,|\,z\,|^2\right) = \tag{4.22}$$

$$= \frac{1}{{}_p\Psi_q\left(\begin{matrix}(\mathrm{a,A})\\(\mathrm{b,B})\end{matrix}\bigg|\,|\,z\,|^2\right)} {}_p\Psi_q\left(\begin{matrix}(\mathrm{a}+\mathrm{A}\cdot s,\mathrm{A})\\(\mathrm{b}+\mathrm{B}\cdot s,\mathrm{B})\end{matrix}\bigg|\,|\,z\,|^2\right)$$

where we used the expression for the $s$ th order derivative of the F-W function, Eq. (P2a).

Finally, it follows that the Mandel parameter has the expression

$$Q_{|z|} = |\,z\,|^2 \left[\frac{{}_p\Psi_q\left(\begin{matrix}(\mathrm{a}+2\mathrm{A},\mathrm{A})\\(\mathrm{b}+2\mathrm{B},\mathrm{B})\end{matrix}\bigg|\,|\,z\,|^2\right)}{{}_p\Psi_q\left(\begin{matrix}(\mathrm{a}+\mathrm{A},\mathrm{A})\\(\mathrm{b}+\mathrm{B},\mathrm{B})\end{matrix}\bigg|\,|\,z\,|^2\right)} - \frac{{}_p\Psi_q\left(\begin{matrix}(\mathrm{a}+\mathrm{A},\mathrm{A})\\(\mathrm{b}+\mathrm{B},\mathrm{B})\end{matrix}\bigg|\,|\,z\,|^2\right)}{{}_p\Psi_q\left(\begin{matrix}(\mathrm{a,A})\\(\mathrm{b,B})\end{matrix}\bigg|\,|\,z\,|^2\right)}\right] \tag{4.23}$$

for every $z \in C \backslash (0)$.

As can be seen, the value of Mandel's parameter, respectively its sign (negative, zero or positive) obviously depends on the value of the set of parameters $(\mathrm{a,A})$, $(\mathrm{b,B})$, as well as the value of the variable $|\,z\,|^2$. Depending on the values of the expression between the right brackets, we can have the following situations:

| $Q_{|z|}$ | The states $\mid z >$ obey the statistics distribution | Which correspond to the nature of system |
|---|---|---|
| $< 0$ | sub-Poissonian | nonclassical states |
| $= 0$ | Poissonian | coherent light |
| $> 0$ | super-Poissonian | thermal states |

Generally, the probability density of the transition from the state $\mid n >$ to state $\mid z >$ or, in other words, the probability that the Fox-Wright coherent state $\mid z >$ coincides with $n$ excitations, i.e. the state $\mid n >$ of the Fock basis, is

$$P_{n;p,q}\left(\mid z \mid^2\right)=\mid < n \mid z >\mid^2 = \frac{1}{{}_p\Psi_q\left(\left.\begin{pmatrix}(a,A)\\(b,B)\end{pmatrix}\right|\mid z \mid^2\right)} \frac{\prod_{i=1}^{p}\Gamma\left(a_i + A_i\, n\right)}{\prod_{j=1}^{q}\Gamma\left(b_j + B_j\, n\right)} \frac{\left(\mid z \mid^2\right)^n}{n!} \tag{4.24}$$

We recall that the Poissonian probability density has the expression

$$P_{n;p,q}^{\text{Poiss}}\left(\mid z \mid^2\right)=\exp\left(-\mid z \mid^2\right)\frac{\left(\mid z \mid^2\right)^n}{n!} \tag{4.25}$$

so that the following inequality is valid $P_{n;p,q}^{\text{sub-Poiss}} < P_{n;p,q}^{\text{Poiss}} < P_{n;p,q}^{\text{super-Poiss}}$.

**e.) *The set of CSs must be temporally stable, in sense that***

$$e^{-\mathrm{i}\frac{1}{\hbar}\hat{\mathcal{H}}t}/z>=\mid e^{-\mathrm{i}\omega t}\, z > \tag{4.26}$$

The CCs satisfy this requirement, but let's see how and if the set of F-WCSs can also satisfy it.

$$e^{-\mathrm{i}\frac{1}{\hbar}\hat{\mathcal{H}}t}/z>=\frac{1}{\sqrt{{}_p\Psi_q\left(\left.\begin{pmatrix}(a,A)\\(b,B)\end{pmatrix}\right|\mid z \mid^2\right)}}\sum_{n=0}^{\infty}\frac{z^n}{\sqrt{{}_p\tilde{\rho}_q(n)}}\, e^{-\mathrm{i}\omega t\,{}_p e_q(n)}\mid n > \tag{4.27}$$

Let's transform the complex exponential $e^{-\mathrm{i}\omega t\,{}_p e_q(n)}$ that is, let's develop ${}_p e_q(n)$ in a Power series with respect to $n$.

$$\begin{aligned}{}_p e_q(n)&=n\,{}_p f_q(n) = n\,\frac{\prod_{j=1}^{q}\left(b_j + B_j\cdot(n-1)\right)}{\prod_{i=1}^{p}\left(a_i + A_i\cdot(n-1)\right)}\equiv {}_p f_q(0)\, n + {}_p f_q^{(1)}(0)\, n^2 + O\left(n^3\right)=\\[2mm]&=\frac{\prod_{j=1}^{q}\left(b_j - B_j\right)}{\prod_{i=1}^{p}\left(a_i - A_i\right)}\, n + {}_p f_q^{(1)}(0)\, n^2 + O\left(n^3\right)\end{aligned} \tag{4.28}$$

If we use only the first-order approximation, that is

$$
{}_p e_q(n) \approx {}_p f_q(0)\, n = {}_p f_q(0)\, n = \frac{\prod\limits_{j=1}^{q}\left(b_j - B_j\right)}{\prod\limits_{i=1}^{p}\left(a_i - A_i\right)}\, n \tag{4.29}
$$

the expression becomes

$$
e^{-i\frac{1}{\hbar}\hat{\mathcal{H}}t}\,|z> = \frac{1}{\sqrt{{}_p\Psi_q\!\left(\!\begin{array}{c}(a,A)\\(b,B)\end{array}\!\Big|\,|z|^2\right)}}\sum_{n=0}^{\infty}\frac{\left(z\,e^{-i\omega t\, {}_p f_q(0)}\right)^n}{\sqrt{{}_p\widetilde{\rho}_q(n)}}\,|n> \equiv \,|z\,e^{-i\omega t\, {}_p f_q(0)}> \tag{4.30}
$$

$$
e^{-i\frac{1}{\hbar}\hat{\mathcal{H}}t}\,|z> \approx \,|z\,e^{-i\omega t\, {}_p f_q(0)}> \tag{4.31}
$$

It can be said, therefore, that F-WCSs remain coherent in time only if we limit ourselves to the first-order approximation, labeled by the temporal dependent variable $z(t) \equiv z\,e^{-i\omega t\, {}_p f_q(0)}$. This is due to the fact that the energy spectrum is nonlinear with respect to the principal quantum number $n$.

*Observation*: The F-WCSs can also be defined in the *Klauder-Perelomov manner*, $|\hat{z}>$ that is, as result of the action of the displacement operator on the vacuum state [23]. For the distinction, we will use the notation for the complex variable $z$ by $\widetilde{z}$, and also for other sizes with the tilde (~) symbol.

$$
|\widetilde{z}> = \frac{1}{\sqrt{\widetilde{N}\!\left(|\widetilde{z}|^2\right)}}\exp\!\left(\widetilde{z}\,\hat{\mathcal{A}}_+ - \widetilde{z}^*\,\hat{\mathcal{A}}_-\right)|0> \tag{4.32}
$$

Because, according to the DOOT rules, the operators $\hat{\mathcal{A}}_+$ and $\hat{\mathcal{A}}_-$ are commutable, so that the Baker–Campbell–Hausdorff formula so that the exponential will be written as

$$
\exp\!\left(\widetilde{z}\,\hat{\mathcal{A}}_+ - \widetilde{z}^*\,\hat{\mathcal{A}}_-\right) = \exp\!\left(\widetilde{z}\,\hat{\mathcal{A}}_+\right)\exp\!\left(-\widetilde{z}^*\,\hat{\mathcal{A}}_-\right) \tag{4.33}
$$

Taking into account that the action of the annihilation operator on the vacuum state it does not change it $\hat{\mathcal{A}}_-|0> = |0>$, we can give up the exponential $\exp\!\left(-\widetilde{z}^*\,\hat{\mathcal{A}}_-\right)$, so that CSs of the Klauder-Perelomov manner can also be defined also as

$$
|\widetilde{z}> = \frac{1}{\sqrt{\widetilde{N}\!\left(|\widetilde{z}|^2\right)}}\exp\!\left(\widetilde{z}\,\hat{\mathcal{A}}_+\right)|0> \tag{4.34}
$$

Developing the exponential in power series, we will have, successively

$$
|\widetilde{z}> = \frac{1}{\sqrt{\widetilde{N}\!\left(|\widetilde{z}|^2\right)}}\exp\!\left(\widetilde{z}\,\hat{\mathcal{A}}_+\right)|0> = \frac{1}{\sqrt{\widetilde{N}\!\left(|\widetilde{z}|^2\right)}}\sum_{n=0}^{\infty}\frac{(\widetilde{z})^n}{n!}\left(\hat{\mathcal{A}}_+\right)^n|0> =
$$

$$
= \frac{1}{\sqrt{\widetilde{N}\!\left(|\widetilde{z}|^2\right)}}\sum_{n=0}^{\infty}\sqrt{\frac{{}_p\rho_q(n)}{(n!)^2}}\,(\widetilde{z})^n\,|n> \tag{4.35}
$$

Using the notations

$$
{}_q\widetilde{\rho}_p(n) \equiv \frac{(n!)^2}{{}_p\rho_q(n)}\ ,\quad {}_q\widetilde{\Psi}_p\!\left(\!\begin{array}{c}(b,B)\\(a,A)\end{array}\!\Big|\,|\widetilde{z}|^2\right) \equiv {}_q\widetilde{\Psi}_p\!\left[\begin{array}{c}(b_1,B_1)\ (b_2,B_2)\ ...\ (b_q,B_q)\\(a_1,A_1)\ (a_2,A_2)\ ...\ (a_p,A_p)\end{array};\,|\widetilde{z}|^2\right] \tag{4.36}
$$

the set of F-WCSs, defined in the Klauder-Perelomov manner, has the following final expression

$$|\tilde{z}> = \frac{1}{\sqrt{_q\tilde{\Psi}_p\left(\binom{(b,B)}{(a,A)}\Big|\ |\tilde{z}|^2\right)}}\sum_{n=0}^{\infty}\frac{\tilde{z}^n}{\sqrt{_q\tilde{\rho}_p(n)}}|\ n> = \frac{1}{\sqrt{_q\tilde{\Psi}_p\left(\binom{(b,B)}{(a,A)}\Big|\ |\tilde{z}|^2\right)}}\exp\left(\tilde{z}\ \hat{\mathcal{A}}_{+}\right)|\ 0> \qquad (4.37)$$

Similar to the case of Barut-Girardello CSs, it is easily deduced that the integration measure has the expression

$$d\tilde{\mu}(\tilde{z}) = \tilde{C}_{q,p}(b,B,a,A)\frac{d^2\tilde{z}}{\pi}\ _q\tilde{\Psi}_p\left(\binom{(b,B)}{(a,A)}\Big|\ |\tilde{z}|^2\right)G_{q,p+1}^{p+1,0}\left(\frac{\prod\limits_{j=1}^{q}B_j}{\prod\limits_{i=1}^{p}A_i}|\tilde{z}|^2\ \Bigg|\ \begin{matrix} /\ ; & \left\{\frac{b_j}{B_j}-1\right\}_1^q \\ 0\ ,\ \left\{\frac{a_i}{A_i}-1\right\}_1^p\ ; & / \end{matrix}\right)$$

$$(4.38)$$

$$\tilde{C}_{q,p}(b,B,a,A) \equiv \frac{\prod\limits_{j=1}^{q}B_j}{\prod\limits_{i=1}^{p}A_i}\ \frac{\prod\limits_{i=1}^{p}\Gamma(a_i)}{\prod\limits_{i=1}^{p}\Gamma\left(\frac{a_i}{A_i}\right)}\ \frac{\prod\limits_{j=1}^{q}\Gamma\left(\frac{b_j}{B_j}\right)}{\prod\limits_{j=1}^{q}\Gamma(b_j)} \qquad (4.39)$$

On the other hand, the unity decomposition relation for Klauder-Perelomov CSs will be written as

$$I = \int d\tilde{\mu}(\tilde{z})|\ \tilde{z}><\tilde{z}| = \frac{\prod\limits_{j=1}^{q}\Gamma(b_j)}{\prod\limits_{i=1}^{p}\Gamma(a_i)}\int d\tilde{\mu}(\tilde{z})\frac{1}{_q\tilde{\Psi}_p\left(\binom{(b,B)}{(a,A)}\Big|\ |\tilde{z}|^2\right)}\#\exp\left(\tilde{z}\ \hat{\mathcal{A}}_{+}\right)|\ 0><0|\exp\left(\tilde{z}^*\ \hat{\mathcal{A}}_{-}\right)\# \qquad (4.40)$$

Using Eq. (3.22), finally we obtain the following integral in complex space

$$\int\frac{d^2\tilde{z}}{\pi}G_{q,p+1}^{p+1,0}\left(\frac{\prod\limits_{j=1}^{q}B_j}{\prod\limits_{i=1}^{p}A_i}|\tilde{z}|^2\ \Bigg|\ \begin{matrix} /\ ; & \left\{\frac{b_j}{B_j}-1\right\}_1^q \\ 0\ ,\ \left\{\frac{a_i}{A_i}-1\right\}_1^p\ ; & / \end{matrix}\right)\#\exp\left(\tilde{z}\ \hat{\mathcal{A}}_{+}\right)\exp\left(\tilde{z}^*\ \hat{\mathcal{A}}_{-}\right)\# =$$

$$= \tilde{C}_{q,p}(b,B,a,A)\ _p\Psi_q\left(\binom{(a,A)}{(b,B)}\Big|\ \hat{\mathcal{A}}_{+}\hat{\mathcal{A}}_{-}\right) \qquad (4.41)$$

This relation is, in fact, a new integral representation of the F-W function.

From the algebraic point of view, it is observed that F-WCSs, defined in the Klauder Perelomov manner, have the same mathematical structure as F-WCSs, defined in the Barut-Girardello manner. The difference lies in the fact that the place of the indices $p$ and $q$, as well as the sets of numbers $\left\{(a_i,A_i)_1^p\right\}$ and $\left\{(b_j,B_j)_1^q\right\}$ have been interchanged. This interchange will be reflected in all expressions that will refer to these types of coherent states. This is an expression of the dualism of the two definitions of CSs (see, e.g. [26] and references therein).

Let us now examine the harmonic limit of F-WCSs. This is achieved when the modulus of the complex numbers $A_i$ and $B_j$ are is very small, tending to zero, $|A_i| \to 0$ and $|B_j| \to 0$, and the numbers $a_i$ and $b_j$ are finite, so their ratios are very large: $\left|\dfrac{a_i}{A_i}\right| \to \infty$ and $\left|\dfrac{b_j}{B_j}\right| \to \infty$ .

We will use the well-known limit (formula 8.328.2, pp. 895 from [27])

$$\lim_{|\varsigma| \to \infty} \frac{\Gamma(\varsigma + n)}{\Gamma(\varsigma)} e^{-n \ln \varsigma} = 1 \tag{4.42}$$

it follows that (see, Appendix D from [27])

$$\Gamma(\varsigma + n) \approx \Gamma(\varsigma) \varsigma^n \quad , \quad |\varsigma| \to \infty \tag{4.43}$$

Equation (1.4) leads us to the expression

$$\Gamma(a_i + A_i n) = A_i^n \frac{\Gamma\left(\dfrac{a_i}{A_i} + n\right)}{\Gamma\left(\dfrac{a_i}{A_i}\right)} \Gamma(a_i) \approx \Gamma(a_i) A_i^n \tag{4.44}$$

Consequently, the limit of the structure constants is

$$\lim_{\substack{\left|\frac{a_i}{A_i}\right| \to \infty \\ \left|\frac{b_j}{B_j}\right| \to \infty}} \rho\left(n \left|\begin{matrix}(a,A) \\ (b,B)\end{matrix}\right.\right) = n! \lim_{\substack{\left|\frac{a_i}{A_i}\right| \to \infty \\ \left|\frac{b_j}{B_j}\right| \to \infty}} \frac{\prod_{j=1}^{q} \Gamma(b_j + B_j n)}{\prod_{i=1}^{p} \Gamma(a_i + A_i n)} \approx n! \frac{\prod_{j=1}^{q} \Gamma(b_j)}{\prod_{i=1}^{p} \Gamma(a_i)} \left(\frac{\prod_{j=1}^{q} B_j}{\prod_{i=1}^{p} A_i}\right)^n \tag{4.45}$$

respectively, the F-W function, Eq. (2.1)

$$\lim_{\substack{\left|\frac{a_i}{A_i}\right| \to \infty \\ \left|\frac{b_j}{B_j}\right| \to \infty}} {}_p\Psi_q\left(\begin{matrix}(a,A) \\ (b,B)\end{matrix}\middle| |z|^2\right) = \sum_{n=0}^{\infty} \frac{(|z|^2)^n}{\lim_{\substack{\left|\frac{a_i}{A_i}\right| \to \infty \\ \left|\frac{b_j}{B_j}\right| \to \infty}} {}_p\rho_q(n)} = \frac{\prod_{i=1}^{p} \Gamma(a_i)}{\prod_{j=1}^{q} \Gamma(b_j)} \sum_{n=0}^{\infty} \frac{1}{n!}\left(\frac{\prod_{i=1}^{p} A_i}{\prod_{j=1}^{q} B_j}|z|^2\right)^n = \tag{4.46}$$

$$= \frac{\prod_{i=1}^{p} \Gamma(a_i)}{\prod_{j=1}^{q} \Gamma(b_j)} \exp\left(\frac{\prod_{i=1}^{p} A_i}{\prod_{j=1}^{q} B_j}|z|^2\right)$$

Finally, the harmonic limit of the F-WCSs becomes, see also Eq. (3.15)

$$\lim_{\substack{\left|\frac{a_i}{A_i}\right|\to\infty \\ \left|\frac{b_j}{B_j}\right|\to\infty}} |z> = \frac{1}{\sqrt{\lim_{\substack{\left|\frac{a_i}{A_i}\right|\to\infty \\ \left|\frac{b_j}{B_j}\right|\to\infty}} {}_p\Psi_q\left(\begin{matrix}(a,A)\\(b,B)\end{matrix}\Big|\,|z|^2\right)}} \sum_{n=0}^{\infty} \frac{z^n}{\sqrt{\lim_{\substack{\left|\frac{a_i}{A_i}\right|\to\infty \\ \left|\frac{b_j}{B_j}\right|\to\infty}} {}_p\rho_q(n)}} |n> =$$

(4.47)

$$= \exp\left(-\frac{\prod_{i=1}^{p}A_i}{\prod_{j=1}^{q}B_j}\,|z|^2\right)\sum_{n=0}^{\infty}\sqrt{\left(\frac{\prod_{i=1}^{p}A_i}{\prod_{j=1}^{q}B_j}\right)^n}\frac{z^n}{\sqrt{n!}}|n>$$

In the particular case, if $A_i = B_j = 1$, the canonical CSs expression is obtained

$$|z> = \exp\left(-|z|^2\right)\sum_{n=0}^{\infty}\frac{z^n}{\sqrt{n!}}|n>$$

(4.48)

## 5. Fox-Wright coherent states involved in thermal states

The states of a quantum system at thermodynamic equilibrium with the external environment (called the "bath") at temperature $T$ are mixed states, and are characterized by the canonical density operator

$$\rho(\beta) = \frac{1}{Z(\beta)}\sum_{n=0}^{\infty}\exp\left(-\beta E_n\right)|n><n|$$

(5.1)

where $\beta = (k_B T)^{-1}$ is the temperature parameter and $k_B$ is the Boltzmann's constant.

Using the expressions of the action of the operators $\hat{\mathcal{A}}_-$ and $\hat{\mathcal{A}}_+$ on the vacuum state and taking into account the rules of the DOOT technique and Eq. (3.5), the above expression is written as

$$\rho(\beta) = \frac{1}{Z(\beta)}\sum_{n=0}^{\infty}\exp\left(-\beta E_n\right)\frac{\#\left(\hat{\mathcal{A}}_+\hat{\mathcal{A}}_-\right)^n\#}{{}_p g_q(n)}|0><0| =$$

$$= \frac{1}{Z(\beta)}\#\frac{1}{{}_p\Psi_q\left(\begin{matrix}(a,A)\\(b,B)\end{matrix}\Big|\,\hat{\mathcal{A}}_+\hat{\mathcal{A}}_-\right)}\#\sum_{n=0}^{\infty}\exp\left(-\beta E_n\right)\frac{\#\left(\hat{\mathcal{A}}_+\hat{\mathcal{A}}_-\right)^n\#}{{}_p\rho_q(n)}$$

(5.2)

where the partition function $Z(\beta)$ is obtained as a consequence of the normalization of density operator to unity $\mathrm{Tr}\rho = 1$.

We will do the calculation in the F-WCSs representation, using Eq. (4.10):

$$\mathrm{Tr}\,\rho = \int d\mu(z) <z\,|\,\rho\,|\,z> = \frac{1}{Z(\beta)} \sum_{n=0}^{\infty} \frac{\exp(-\beta E_n)}{_p\rho_q(n)} \int d\mu(z) <z\,|\# \frac{\left(\hat{\mathcal{A}}_+ \hat{\mathcal{A}}_-\right)^n}{_p\Psi_q\left(\begin{matrix}(a,A)\\(b,B)\end{matrix}\bigg|\,\hat{\mathcal{A}}_+ \hat{\mathcal{A}}_-\right)}\#|\,z> =$$

$$= \frac{C_{p,q}(a,A,b,B)}{Z(\beta)} \sum_{n=0}^{\infty} \frac{\exp(-\beta E_n)}{_p\rho_q(n)} \int_0^{\infty} d\left(|z|^2\right) G_{p,q+1}^{q+1,0}\left(\begin{matrix} \prod_{i=1}^{p} A_i \\ \prod_{j=1}^{q} B_j \end{matrix}|z|^2 \,\Bigg|\, \begin{matrix} / \;; & \left\{\frac{a_i}{A_i}-1\right\}_1^p \\ 0,\left\{\frac{b_j}{B_j}-1\right\}_1^q \;; & / \end{matrix}\right)\left(|z|^2\right)^n =$$

$$= \frac{C_{p,q}(a,A,b,B)}{Z(\beta)} \sum_{n=0}^{\infty} \frac{\exp(-\beta E_n)}{_p\rho_q(n)} \frac{1}{C_{p,q}(a,A,b,B)} {}_p\rho_q(n) = \frac{1}{Z(\beta)} \sum_{n=0}^{\infty} e^{-\beta E_n} = 1$$

(5.3)

Consequently, the partition function is

$$Z(\beta) = \sum_{n=0}^{\infty} \exp(-\beta E_n) \tag{5.4}$$

Let's calculate the trace of the density operator using the F-WCSs representation.

$$\rho(\beta) = \frac{1}{Z(\beta)} \sum_{n=0}^{\infty} \exp(-\beta E_n) \frac{\#\left(\hat{\mathcal{A}}_+ \hat{\mathcal{A}}_-\right)^n \#}{_p g_q(n)} |\,0><0\,| =$$

$$= \frac{1}{Z(\beta)} \# \frac{1}{_p\Psi_q\left(\begin{matrix}(a,A)\\(b,B)\end{matrix}\bigg|\,\hat{\mathcal{A}}_+ \hat{\mathcal{A}}_-\right)} \# \sum_{n=0}^{\infty} \exp(-\beta E_n) \frac{\#\left(\hat{\mathcal{A}}_+ \hat{\mathcal{A}}_-\right)^n \#}{_p\rho_q(n)}$$

(5.5)

The above expression simplifies for quantum systems that have a linear energy spectrum (linear oscillator, pseudoharmonic oscillator, Landau levels), i.e. $E_n = E_0 + \hbar\omega n$, and for these situations we obtain

$$\rho(\beta) = \left(1 - e^{-\beta\hbar\omega}\right) \# \frac{1}{_p\Psi_q\left(\begin{matrix}(a,A)\\(b,B)\end{matrix}\bigg|\,\hat{\mathcal{A}}_+ \hat{\mathcal{A}}_-\right)} \# \sum_{n=0}^{\infty} \frac{\#\left(e^{-\beta\hbar\omega} \hat{\mathcal{A}}_+ \hat{\mathcal{A}}_-\right)^n \#}{_p\rho_q(n)} =$$

$$= \left(1 - e^{-\beta\hbar\omega}\right) \# \frac{1}{_p\Psi_q\left(\begin{matrix}(a,A)\\(b,B)\end{matrix}\bigg|\,\hat{\mathcal{A}}_+ \hat{\mathcal{A}}_-\right)} {}_p\Psi_q\left(\begin{matrix}(a,A)\\(b,B)\end{matrix}\bigg|\,e^{-\beta\hbar\omega} \hat{\mathcal{A}}_+ \hat{\mathcal{A}}_-\right) \#$$

(5.6)

The diagonal elements of the density operator, in the CSs representation are called the Husimi's distribution function $Q\left(|z|^2\right) \equiv <z\,|\,\rho(\beta)\,|\,z>$ [28]:

$$Q\left(|z|^2\right) = \frac{1}{Z(\beta)} \frac{1}{_p\Psi_q\left(\begin{matrix}(a,A)\\(b,B)\end{matrix}\bigg|\,|z|^2\right)} \sum_{n=0}^{\infty} \exp(-\beta E_n) \frac{\left(|z|^2\right)^n}{_p\rho_q(n)} \tag{5.7}$$

respectively for systems with linear energy spectra

$$Q\left(|z|^2\right) = \left(1 - e^{-\beta\hbar\omega}\right) \frac{1}{{}_p\Psi_q\left(\begin{matrix}(a,A)\\(b,B)\end{matrix}\middle||z|^2\right)} {}_p\Psi_q\left(\begin{matrix}(a,A)\\(b,B)\end{matrix}\middle|e^{-\beta\hbar\omega}|z|^2\right) \qquad (5.8)$$

Like any distribution function, Husimi's distribution function is normalized to unity:

$$\int d\mu(z)Q\left(|z|^2\right) = \int d\mu(z)<z|\rho(\beta)|z> = 1 \qquad (5.9)$$

Moreover, the so called the diagonal expansion of the density operator in the CSs representation is

$$\rho(\beta) = \int d\mu(z)P_{p,q}\left(|z|^2\right)|z><z| \qquad (5.10)$$

Using the expression for the vacuum projector, Eq. (3.2) we have

$$\rho(\beta) = C_{p,q}(a,A,b,B)\int\frac{d^2z}{\pi} G_{p,q+1}^{q+1,0}\left(\frac{\prod_{i=1}^{p}A_i}{\prod_{j=1}^{q}B_j}|z|^2\middle|\begin{matrix} / & ; & \left\{\frac{a_i}{A_i}-1\right\}_1^p \\ 0, \left\{\frac{b_j}{B_j}-1\right\}_1^q & ; & / \end{matrix}\right)P_{p,q}\left(|z|^2\right)\times$$

$$\times\# \frac{{}_p\Psi_q\left(\begin{matrix}(a,A)\\(b,B)\end{matrix}\middle|z\hat{\mathcal{A}}_+\right){}_p\Psi_q\left(\begin{matrix}(a,A)\\(b,B)\end{matrix}\middle|z^*\hat{\mathcal{A}}_-\right)}{{}_p\Psi_q\left(\begin{matrix}(a,A)\\(b,B)\end{matrix}\middle|\hat{\mathcal{A}}_+\hat{\mathcal{A}}_-\right)}\#$$

$$(5.11)$$

The angular integral is

$$\int_0^{2\pi}\frac{d\varphi}{2\pi}\#{}_p\Psi_q\left(\begin{matrix}(a,A)\\(b,B)\end{matrix}\middle|z\hat{\mathcal{A}}_+\right){}_p\Psi_q\left(\begin{matrix}(a,A)\\(b,B)\end{matrix}\middle|z^*\hat{\mathcal{A}}_-\right)\# = \sum_{n=0}^{\infty}\frac{\#\left(\hat{\mathcal{A}}_+\hat{\mathcal{A}}_-\right)^n\#}{\left[{}_p\rho_q(n)\right]^2}\left(|z|^2\right)^n \qquad (5.12)$$

and so that we have

$$\rho(\beta) = C_{p,q}(a,A,b,B)\#\frac{1}{{}_p\Psi_q\left(\begin{matrix}(a,A)\\(b,B)\end{matrix}\middle|\hat{\mathcal{A}}_+\hat{\mathcal{A}}_-\right)}\#\sum_{n=0}^{\infty}\frac{\#\left(\hat{\mathcal{A}}_+\hat{\mathcal{A}}_-\right)^n\#}{\left[{}_p\rho_q(n)\right]^2}\times$$

$$\times\int_0^{\infty}d\left(|z|^2\right)G_{p,q+1}^{q+1,0}\left(\frac{\prod_{i=1}^{p}A_i}{\prod_{j=1}^{q}B_j}|z|^2\middle|\begin{matrix} / & ; & \left\{\frac{a_i}{A_i}-1\right\}_1^p \\ 0, \left\{\frac{b_j}{B_j}-1\right\}_1^q & ; & / \end{matrix}\right)P_{p,q}\left(|z|^2\right)\left(|z|^2\right)^n$$

$$(5.13)$$

To have equality, that is, to reach expression (6.2), the integral over the variable $|z|^2$ must be

$$\int_0^\infty d\left(|z|^2\right) G_{p,q+1}^{q+1,0}\left(\begin{matrix}\prod\limits_{i=1}^{p}A_i \\ \prod\limits_{j=1}^{q}B_j\end{matrix}|z|^2 \left| \begin{matrix} & / \; ; & \left\{\dfrac{a_i}{A_i}-1\right\}_1^p \\ 0 \, , \left\{\dfrac{b_j}{B_j}-1\right\}_1^q \; ; & / \end{matrix}\right.\right) P_{p,q}\left(|z|^2\right)\left(|z|^2\right)^n =$$

$$= \frac{1}{Z(\beta)} \frac{\prod\limits_{j=1}^{q}\Gamma\left(\dfrac{b_j}{B_j}\right)}{\prod\limits_{i=1}^{p}\Gamma\left(\dfrac{a_i}{A_i}\right)} \frac{\prod\limits_{j=1}^{q}B_j}{\prod\limits_{i=1}^{p}A_i} \sum_{n=0}^{\infty}\exp\left(-\beta E_n\right) {}_p\rho_q(n) =$$

<div align="right">(5.14)</div>

$$\rho(\beta) = \frac{1}{Z(\beta)}\sum_{n=0}^{\infty}\exp\left(-\beta E_n\right)\frac{\#\left(\hat{\mathcal{A}}_+\hat{\mathcal{A}}_-\right)^n\#}{{}_pg_q(n)}|0><0| =$$

$$= \frac{1}{Z(\beta)}\#\frac{1}{{}_p\Psi_q\left(\begin{matrix}(a,A)\\(b,B)\end{matrix}\middle|\hat{\mathcal{A}}_+\hat{\mathcal{A}}_-\right)}\#\sum_{n=0}^{\infty}\exp\left(-\beta E_n\right)\frac{\#\left(\hat{\mathcal{A}}_+\hat{\mathcal{A}}_-\right)^n\#}{{}_pg_q(n)}$$

<div align="right">(5.15)</div>

## 6. Mathematical feedback of Fox-Wright coherent states

Apart from the applications of F-WCSs in quantum physics (and in particular, in quantum optics), the formalism of these coherent states also leads to a series of mathematical relations and equalities, relating to F-W functions. This fact can be seen as a true feedback from physics (through the coherent states formalism) to mathematics (in the theory of special functions).

At the end of this paper we will only refer to some integrals in which the F-W functions are involved, without excluding the fact that there may be other such feedbacks.

Recall that among the rules of the DOOT technique is that the creation and annihilation operators are treated as simple *c*-numbers. This means that in mathematical relations they can be replaced by simple numbers (or letters), the respective relations retaining its validity. In this regard, we will use the following substitutions: $\hat{\mathcal{A}}_+ \to \lambda$ and $\hat{\mathcal{A}}_- \to \varepsilon$. The numbers $(a, A)$, $(b, B)$ can also be replaced by other scalar quantities.

### a) Integrals involving a product between Meijer's G function and one Fox-Wright function

This type of integral in real space are obtained using Eq. (4.10).

$$Int_1 \equiv \int_0^\infty d\left(|z|^2\right) G_{p,q+1}^{q+1,0}\left(\left.\frac{\prod_{i=1}^p A_i}{\prod_{j=1}^q B_j}\,|z|^2\,\right|\begin{array}{c} / \;\; ; \;\; \left\{\frac{a_i}{A_i}-1\right\}_1^p \\ 0\,,\; \left\{\frac{b_j}{B_j}-1\right\}_1^q \;\; ; \;\; / \end{array}\right)\, {}_r\Psi_l\left(\left.\begin{array}{c}(c,C) \\ (d,D)\end{array}\right|\lambda\,|z|^2\right)=$$

$$=\frac{\prod_{i=1}^p A_i}{\prod_{j=1}^q B_j}\,\frac{\prod_{i=1}^p \Gamma\left(\frac{a_i}{A_i}\right)}{\prod_{i=1}^p \Gamma(a_i)}\,\frac{\prod_{j=1}^q \Gamma(b_j)}{\prod_{j=1}^q \Gamma\left(\frac{b_j}{B_j}\right)}\; {}_{q+1+r}\Psi_{p+l}\left(\left.\begin{array}{c}(b,B),(1,1),(c,C) \\ (a,A),(d,D)\end{array}\right|\lambda\right)= \qquad (6.1)$$

$$=C_{p,q}(a,A,b,B)\; {}_{q+1+r}\Psi_{p+l}\left(\left.\begin{array}{c}(b,B),(1,1),(c,C) \\ (a,A),(d,D)\end{array}\right|\lambda\right)$$

**Proof**

$$Int_1 = \sum_{n=0}^\infty \frac{\prod_{i=1}^r \Gamma(c_i+C_i\,n)}{\prod_{j=1}^l \Gamma(d_j+D_j\,n)}\,\frac{\lambda^n}{n!}\int_0^n d\left(|z|^2\right) G_{p,q+1}^{q+1,0}\left(\left.\frac{\prod_{i=1}^p A_i}{\prod_{j=1}^q B_j}\,|z|^2\,\right|\begin{array}{c} / \;\; ; \;\; \left\{\frac{a_i}{A_i}-1\right\}_1^p \\ 0\,,\; \left\{\frac{b_j}{B_j}-1\right\}_1^q \;\; ; \;\; / \end{array}\right)\left(|z|^2\right)^n=$$

$$=\frac{1}{C_{p,q}(a,A,b,B)}\sum_{n=0}^\infty \frac{\prod_{i=1}^r \Gamma(c_i+C_i\,n)}{\prod_{j=1}^l \Gamma(d_j+D_j\,n)}\,\frac{\lambda^n}{n!}\;{}_p\rho_q(n)=$$

$$=\frac{1}{C_{p,q}(a,A,b,B)}\sum_{n=0}^\infty \frac{\prod_{i=1}^r \Gamma(c_i+C_i\,n)}{\prod_{j=1}^l \Gamma(d_j+D_j\,n)}\,\frac{\lambda^n}{n!}\,\frac{\prod_{j=1}^q \Gamma(b_j+B_j\,n)}{\prod_{i=1}^p \Gamma(a_i+A_i\,n)}=$$

$$=\frac{1}{C_{p,q}(a,A,b,B)}\;{}_{q+1+r}\Psi_{p+l}\left(\left.\begin{array}{c}(b,B),(c,C),\,(1,1), \\ (a,A),\,(d,D)\end{array}\right|\lambda\right)$$

$$\hfill (6.2)$$

where $n!=\Gamma(1+1\cdot n)\equiv(1,1)$. So, we arrived at the announce formula.

  *Particular case*. If we choose $p=q=0$, then $A_i=B_j=1$, and $\lambda=\frac{1}{s}$, $|z|^2=s\,x$, we

have $G_{0,1}^{1,0}(s\,x|0)=\exp(-s\,x)$ and the integral becomes

$$\mathcal{L}\left\{{}_0\Psi_0\right\}(s)=\int_0^\infty dx\, e^{-s\,x}\,{}_r\Psi_l\left(\left.\begin{array}{c}(c,C) \\ (d,D)\end{array}\right|x\right)=\frac{1}{s}\,{}_{r+1}\Psi_l\left(\left.\begin{array}{c}(c,C),(1,1) \\ (d,D)\end{array}\right|\frac{1}{s}\right) \qquad (6.3)$$

and we arrive at the Laplace transform, as in Eq. (2.14).

**b) Integrals involving a product between Meijer's G function and two Fox-Wright functions**

This type of integral results if we starting from Eqs. (4.12) or (3.23).

$$
Int_2 \equiv \int \frac{d^2 z}{\pi} G_{p,q+1}^{q+1,0}\left( \frac{\prod\limits_{i=1}^{p} A_i}{\prod\limits_{j=1}^{q} B_j} |z|^2 \left| \begin{array}{ccc} & / \; ; & \left\{ \frac{a_i}{A_i}-1 \right\}_1^l \\ 0 & , \; \left\{ \frac{b_j}{B_j}-1 \right\}_1^q \; ; & / \end{array} \right. \right)_r \Psi_l\left( \frac{(c,C)}{(d,D)} \middle| \lambda z \right)_\mu \Psi_\nu\left( \frac{(e,E)}{(f,F)} \middle| \varepsilon z^* \right) =
$$

$$
= \frac{1}{C_{p,q}(a,A,b,B)} {}_{q+r+\mu}\Psi_{p+l+\nu}\left( \begin{array}{c} (b,B),\ (c,C),\ (e,E) \\ (a,A),\ (d,D),(f,F) \end{array} \middle| \lambda\varepsilon \right)
$$

(6.4)

**Proof.** We use Eq. (6.5a) and (4.10)

$$
Int_2 = \int\limits_{0}^{\infty} d\left(|z|^2\right) G_{p,q+1}^{q+1,0}\left( \frac{\prod\limits_{i=1}^{p} A_i}{\prod\limits_{j=1}^{q} B_j} |z|^2 \left| \begin{array}{ccc} & / \; ; & \left\{ \frac{a_i}{A_i}-1 \right\}_1^p \\ 0 & , \; \left\{ \frac{b_j}{B_j}-1 \right\}_1^q \; ; & / \end{array} \right. \right) \times
$$

$$
\times \int\limits_{0}^{2\pi} \frac{d\varphi}{2\pi} {}_r\Psi_l\left( \frac{(c,C)}{(d,D)} \middle| \lambda z \right)_\mu \Psi_\nu\left( \frac{(e,E)}{(f,F)} \middle| \varepsilon z^* \right) =
$$

(6.5)

$$
\int\limits_{0}^{2\pi} \frac{d\varphi}{2\pi} {}_r\Psi_l\left( \frac{(c,C)}{(d,D)} \middle| \lambda z \right)_\mu \Psi_\nu\left( \frac{(e,E)}{(f,F)} \middle| \varepsilon z^* \right) =
$$

$$
= \sum\limits_{n=0}^{\infty} \frac{\prod\limits_{i=1}^{r}\Gamma(c_i+C_i\,n)}{\prod\limits_{j=1}^{l}\Gamma(d_l+D_l\,n)} \frac{\lambda^n}{n!} \sum\limits_{m=0}^{\infty} \frac{\prod\limits_{i=1}^{\mu}\Gamma(e_i+E_i\,m)}{\prod\limits_{j=1}^{\nu}\Gamma(f_j+F_j\,m)} \frac{\varepsilon^m}{m!} \int\limits_{0}^{2\pi}\frac{d\varphi}{2\pi} z^n (z^*)^m =
$$

$$
= \sum\limits_{n=0}^{\infty} \frac{\prod\limits_{i=1}^{r}\Gamma(c_i+C_i\,n)}{\prod\limits_{j=1}^{l}\Gamma(d_l+D_l\,n)} \frac{\prod\limits_{i=1}^{\mu}\Gamma(e_i+E_i\,n)}{\prod\limits_{j=1}^{\nu}\Gamma(f_j+F_j\,n)} \frac{(\lambda\varepsilon)^n}{(n!)^2} \left(|z|^2\right)^n
$$

$$Int_2 = \sum_{n=0}^{\infty} \frac{(\lambda\varepsilon)^n}{_r\rho_l(n)\,_\mu\rho_\nu(n)} \int_0^\infty d\left(\mid z\mid^2\right) G_{p,q+1}^{q+1,0}\left(\frac{\prod_{i=1}^p A_i}{\prod_{j=1}^q B_j}\mid z\mid^2 \left|\begin{array}{ccc} / & ; & \left\{\frac{a_i}{A_i}-1\right\}_1^p \\ 0 & , \left\{\frac{b_j}{B_j}-1\right\}_1^q & ; & / \end{array}\right.\right)\left(\mid z\mid^2\right)^n =$$

$$= \frac{1}{C_{p,q}(\mathbf{a,A,b,B})} \sum_{n=0}^{\infty} \frac{\prod_{j=1}^q \Gamma\left(b_j+B_j\cdot n\right)\prod_{j=1}^r \Gamma\left(c_j+C_j\cdot n\right)\prod_{j=1}^\mu \Gamma\left(e_{jr}+E_j\cdot n\right)}{\prod_{i=1}^p \Gamma\left(a_i+A_i\cdot n\right)\prod_{i=1}^l \Gamma\left(d_i+D_i\cdot n\right)\prod_{i=1}^\nu \Gamma\left(f_i+F_i\cdot n\right)} \frac{(\lambda\varepsilon)^n}{n!} =$$

$$= \frac{1}{C_{p,q}(\mathbf{a,A,b,B})}\ _{q+r+\mu}\Psi_{p+l+\nu}\left(\begin{array}{c}(\mathbf{b,B}),(\mathbf{c,C}),(\mathbf{e,E}) \\ (\mathbf{a,A}),(\mathbf{d,D}),(\mathbf{f,F})\end{array}\middle|\lambda\varepsilon\right)$$

(6.6)

This ends the proof.

***Particular case.***

Let's consider the relation between F-W function and hypergeometric function, Eq. (2.12). So, the $Int_2$ becomes

$$\frac{\prod_{i=1}^r \Gamma(c_i)\prod_{i=1}^\mu \Gamma(e_i)}{\prod_{j=1}^l \Gamma(d_j)\prod_{j=1}^\nu \Gamma(f_j)}\int\frac{d^2z}{\pi} G_{p,q+1}^{q+1,0}\left(\frac{\prod_{i=1}^p A_i}{\prod_{j=1}^q B_j}\mid z\mid^2\left|\begin{array}{ccc} / & ; & \left\{\frac{a_i}{A_i}-1\right\}_1^p \\ 0 & , \left\{\frac{b_j}{B_j}-1\right\}_1^q & ; & / \end{array}\right.\right)\times$$

$$\times\ _rF_l\left(\left\{\frac{c_i}{C_i}\right\}_1^r;\ \left\{\frac{d_j}{D_j}\right\}_1^l;\ \frac{\prod_{i=1}^r C_i}{\prod_{j=1}^l D_j}\lambda z\right)\ _\mu F_\nu\left(\left\{\frac{e_i}{E_i}\right\}_1^\mu;\ \left\{\frac{f_j}{F_j}\right\}_1^\nu;\ \frac{\prod_{i=1}^\mu E_i}{\prod_{j=1}^\nu F_j}\varepsilon\,z*\right)=$$

$$=\ _{q+r+\mu}\Psi_{p+l+\nu}\left(\begin{array}{c}(b,\ B),(c,C),(e,\ E) \\ (a,A),(d,\ D),(f,F)\end{array}\middle|\lambda\varepsilon\right)$$

(6.7)

This is a new integral representation of the F-W function, different from those in [9], and which is a consequence of implementing the coherent states formalism.

**Proof**

$$Int_2 = \frac{\prod_{i=1}^{r}\Gamma(c_i)\prod_{i=1}^{\mu}\Gamma(e_i)}{\prod_{j=1}^{l}\Gamma(d_j)\prod_{j=1}^{\nu}\Gamma(f_j)}\int\frac{d^2z}{\pi}G_{p,q+1}^{q+1,0}\left(\frac{\prod_{i=1}^{p}A_i}{\prod_{j=1}^{q}B_j}|z|^2\left|\begin{array}{c}/\ ;\ \left\{\frac{a_i}{A_i}-1\right\}_1^p\\ 0\ ,\ \left\{\frac{b_j}{B_j}-1\right\}_1^q\ ;\ /\end{array}\right.\right)\times$$

$$\times\,_rF_l\left(\left\{\frac{c_i}{C_i}\right\}_1^r\ ;\ \left\{\frac{d_j}{D_j}\right\}_1^l\ ;\ \frac{\prod_{i=1}^{r}C_i}{\prod_{j=1}^{l}D_j}\lambda\,z\right)\,_\mu F_\nu\left(\left\{\frac{e_i}{E_i}\right\}_1^\mu\ ;\ \left\{\frac{f_j}{F_j}\right\}_1^\nu\ ;\ \frac{\prod_{i=1}^{\mu}E_i}{\prod_{j=1}^{\nu}F_j}\varepsilon\,z*\right)=$$

$$=\frac{\prod_{i=1}^{r}\Gamma(c_i)\prod_{i=1}^{\mu}\Gamma(e_i)}{\prod_{j=1}^{l}\Gamma(d_j)\prod_{j=1}^{\nu}\Gamma(f_j)}\sum_{n=0}^{\infty}\frac{\prod_{i=1}^{r}\left(\frac{c_i}{C_i}\right)_n\prod_{i=1}^{\mu}\left(\frac{e_i}{E_i}\right)_n}{\prod_{j=1}^{l}\left(\frac{d_j}{D_j}\right)_n\prod_{j=1}^{\nu}\left(\frac{f_j}{F_j}\right)_n}\left(\frac{\prod_{i=1}^{r}C_i}{\prod_{j=1}^{l}D_j}\lambda\right)^n\left(\frac{\prod_{i=1}^{\mu}E_i}{\prod_{j=1}^{\nu}F_j}\varepsilon\right)^n\frac{1}{(n!)^2}\times$$

$$\times\int\frac{d^2z}{\pi}G_{p,q+1}^{q+1,0}\left(\frac{\prod_{i=1}^{p}A_i}{\prod_{j=1}^{q}B_j}|z|^2\left|\begin{array}{c}/\ ;\ \left\{\frac{a_i}{A_i}-1\right\}_1^p\\ 0\ ,\ \left\{\frac{b_j}{B_j}-1\right\}_1^q\ ;\ /\end{array}\right.\right)\left(|z|^2\right)^n \tag{6.8}$$

$$\frac{\prod_{i=1}^{r}\left(\frac{c_i}{C_i}\right)_n\prod_{i=1}^{\mu}\left(\frac{e_i}{E_i}\right)_n}{\prod_{j=1}^{r}\left(\frac{d_j}{D_j}\right)_n\prod_{j=1}^{\nu}\left(\frac{f_j}{F_j}\right)_n}=\frac{\prod_{i=1}^{r}\Gamma\left(\frac{c_i}{C_i}+n\right)}{\prod_{i=1}^{r}\Gamma\left(\frac{c_i}{C_i}\right)}\frac{\prod_{j=1}^{r}\Gamma\left(\frac{d_j}{D_j}\right)}{\prod_{j=1}^{r}\Gamma\left(\frac{d_j}{D_j}+n\right)}\frac{\prod_{i=1}^{\mu}\Gamma\left(\frac{e_i}{E_i}+n\right)}{\prod_{i=1}^{\mu}\Gamma\left(\frac{e_i}{E_i}\right)}\frac{\prod_{j=1}^{\nu}\Gamma\left(\frac{f_j}{F_j}\right)}{\prod_{j=1}^{\nu}\Gamma\left(\frac{f_j}{F_j}+n\right)}=$$

$$=\frac{\prod_{j=1}^{l}\Gamma\left(\frac{d_j}{D_j}\right)\prod_{j=1}^{\nu}\Gamma\left(\frac{f_j}{F_j}\right)}{\prod_{i=1}^{r}\Gamma\left(\frac{c_i}{C_i}\right)\prod_{i=1}^{\mu}\Gamma\left(\frac{e_i}{E_i}\right)}\frac{\prod_{i=1}^{r}\Gamma\left(\frac{c_i}{C_i}+n\right)\prod_{i=1}^{\mu}\Gamma\left(\frac{e_i}{E_i}+n\right)}{\prod_{j=1}^{l}\Gamma\left(\frac{d_j}{D_j}+n\right)\prod_{j=1}^{\nu}\Gamma\left(\frac{f_j}{F_j}+n\right)}=$$

$$=\frac{\prod_{j=1}^{l}\Gamma(d_j)\prod_{j=1}^{\nu}\Gamma(f_j)}{\prod_{i=1}^{r}\Gamma(c_i)\prod_{i=1}^{\mu}\Gamma(e_i)}\left(\frac{\prod_{j=1}^{l}d_j\prod_{j=1}^{\nu}f_j}{\prod_{i=1}^{r}c_i\prod_{i=1}^{\mu}e_i}\right)^n\frac{\prod_{i=1}^{r}\Gamma(c_i+C_in)\prod_{i=1}^{\mu}\Gamma(e_i+E_in)}{\prod_{j=1}^{l}\Gamma(d_j+D_jn)\prod_{j=1}^{\nu}\Gamma(f_j+F_jn)} \tag{6.9}$$

According to the Eq. (4.10), the integral is

$$\int_0^\infty d\left(|z|^2\right) G_{p,q+1}^{q+1,0}\left(\begin{array}{c}\prod\limits_{i=1}^{p}A_i\\\prod\limits_{j=1}^{q}B_j\end{array}|z|^2\left|\begin{array}{ccc}/ & ; & \left\{\dfrac{a_i}{A_i}-1\right\}_1^p\\ 0, & \left\{\dfrac{b_j}{B_j}-1\right\}_1^q ; & /\end{array}\right.\right)\left(|z|^2\right)^n = \left(\dfrac{\prod\limits_{j=1}^{q}B_j}{\prod\limits_{i=1}^{p}A_i}\right)^n n! \dfrac{\prod\limits_{j=1}^{q}\Gamma\left(\dfrac{b_j}{B_j}+n\right)}{\prod\limits_{i=1}^{p}\Gamma\left(\dfrac{a_i}{A_i}+n\right)} =$$

$$= n! \dfrac{\prod\limits_{i=1}^{p}\Gamma(a_i) \prod\limits_{j=1}^{q}\Gamma\left(\dfrac{b_j}{B_j}\right) \prod\limits_{j=1}^{q}\Gamma(b_j+B_j\,n)}{\prod\limits_{i=1}^{p}\Gamma\left(\dfrac{a_i}{A_i}\right) \prod\limits_{j=1}^{q}\Gamma(b_j) \prod\limits_{i=1}^{p}\Gamma(a_i+A_i\,n)}$$

$$(6.10)$$

$$\text{Int}_2 = \dfrac{\prod\limits_{i=1}^{r}\Gamma(c_i)}{\prod\limits_{i=1}^{r}\Gamma\left(\dfrac{c_i}{C_i}\right)} \dfrac{\prod\limits_{j=1}^{l}\Gamma\left(\dfrac{d_j}{D_j}\right)}{\prod\limits_{j=1}^{l}\Gamma(d_j)} \dfrac{\prod\limits_{i=1}^{\mu}\Gamma(e_i)}{\prod\limits_{i=1}^{\mu}\Gamma\left(\dfrac{e_i}{E_i}\right)} \dfrac{\prod\limits_{j=1}^{\nu}\Gamma\left(\dfrac{f_j}{F_j}\right)}{\prod\limits_{j=1}^{\nu}\Gamma(f_j)} \times$$

$$\times \sum_{n=0}^{\infty} \dfrac{\prod\limits_{j=1}^{q}\Gamma(b_j+B_j\,n)}{\prod\limits_{i=1}^{p}\Gamma(a_i+A_i\,n)} \dfrac{\prod\limits_{i=1}^{r}\Gamma\left(\dfrac{c_i}{C_i}+n\right)}{\prod\limits_{j=1}^{l}\Gamma\left(\dfrac{d_j}{D_j}+n\right)} \dfrac{\prod\limits_{i=1}^{\mu}\Gamma\left(\dfrac{e_i}{E_i}+n\right)}{\prod\limits_{j=1}^{\nu}\Gamma\left(\dfrac{f_j}{F_j}+n\right)} \dfrac{1}{n!}\left(\dfrac{\prod\limits_{i=1}^{r}C_i \prod\limits_{i=1}^{\mu}E_i}{\prod\limits_{j=1}^{l}D_j \prod\limits_{j=1}^{\nu}F_j}\lambda\varepsilon\right)^n =$$

$$= \sum_{n=0}^{\infty} \dfrac{\prod\limits_{j=1}^{q}\Gamma(b_j+B_j\,n)}{\prod\limits_{i=1}^{p}\Gamma(a_i+A_i\,n)} \dfrac{\prod\limits_{i=1}^{r}\Gamma(c_i+C_i\,n)}{\prod\limits_{j=1}^{l}\Gamma(d_j+D_j\,n)} \dfrac{\prod\limits_{i=1}^{\mu}\Gamma(e_i+E_i\,n)}{\prod\limits_{j=1}^{\nu}\Gamma(f_j+F_j\,n)} \dfrac{(\lambda\varepsilon)^n}{n!} =$$

$$= {}_{q+r+\mu}\Psi_{p+l+\nu}\left(\begin{array}{c}(b,\,B),(c,\,C),(e,\,E)\\(a,\,A),(d,\,D),(f,\,F)\end{array}\right|\lambda\varepsilon\right)$$

$$(6.11)$$

This ended the poof.

### c) Integrals involving Meijer G-function and exponentials arising from coherent states of Klauder-Perelomov type

Considering the Eq. (4.41) and taking into account that the DOOT formalism treats the creation and annihilation operators as c-numbers, so it allows their replacement by constants, we will make the replacements: $\hat{\mathcal{A}}_+ \to \lambda$ and $\hat{\mathcal{A}}_- \to \varepsilon$.

Another new integral representation of the F-W functions then becomes from using the coherent states of the Klauder-Perelomov kind, Eq. (4.41):

$$\int \frac{d^2\tilde{z}}{\pi} G_{q,p+1}^{p+1,0} \left( \begin{array}{c} \prod\limits_{j=1}^{q} B_j \\ \prod\limits_{i=1}^{p} A_i \end{array} \, |\tilde{z}|^2 \, \left| \begin{array}{ccc} & / \;\; ; & \left\{ \dfrac{b_j}{B_j} - 1 \right\}_1^q \\ 0 \, , & \left\{ \dfrac{a_i}{A_i} - 1 \right\}_1^p \; ; & / \end{array} \right. \right) \# \exp\left( \lambda \tilde{z} \right) \exp\left( \varepsilon \tilde{z}^* \right) \# = \tag{6.12}$$

$$= \tilde{C}_{q,p}\left( b, B, a, A \right) {}_p\Psi_q \left( \left. \begin{array}{c} (a, A) \\ (b, B) \end{array} \right| \lambda\,\varepsilon \right)$$

***Particular case.***

If $p = q = 0$ (and $A_i = B_j = 1$ ), we obtain an integral often used in quantum optics [21].

$$\int \frac{d^2\tilde{z}}{\pi} e^{-|\tilde{z}|^2} \, \exp\left( \lambda \tilde{z} \right) \exp\left( \varepsilon \tilde{z}^* \right) = e^{\lambda\,\varepsilon} \tag{6.13}$$

which can also be written in the form

$$\int \frac{d^2\tilde{z}}{\pi} \exp\left[ -(\tilde{z} - \lambda)(\tilde{z}^\bullet - \varepsilon) \right] = 1 \tag{6.14}$$

## 7. Concluding remarks

In the paper, which has an interdisciplinary character, therefore in the field of mathematical physics, we focused our attention on a new application of Fox-Wright functions: the coherent states formalism. This is an important tool in quantum mechanics, specifically in quantum optics. We constructed the Barut-Girardello type coherent states, which have the Fox-Wright function as their normalization function. Therefore, these coherent states can be called Fox-Wright coherent states. We have demonstrated that these coherent states satisfy all the requirements imposed on states of this type (the so-called "Klauder's minimal prescriptions").

We also applied the Fox-Wrifgh coherent states formalism to mixed (thermal) states, characterized by the density operator, deducing the two types of distributions associated with it (Husimi's distribution Q and diagonal P distribution).

This approach broadens the area of applicability in physics, therefore in a non-mathematical field, of Fox-Wright functions. Conversely, the Fox-Wright coherent states formalism allows, as a true feedback, the obtaining of new integrals involving the Fox-Wright functions, thus enriching their properties, as well as new integral representations of the Fox-Wright functions.